\documentclass[aps,prd,amsfonts,amssymb, amsmath, showpacs, showkeys, a4paper, nofootinbib, superscriptaddress, 11pt, raggedbottom]{revtex4-2}

\usepackage{braket}
\usepackage{bm}
\usepackage{graphicx}
\usepackage{slashed}
\usepackage{subfig}
\usepackage{mathrsfs}
\usepackage{color}
\usepackage{mathtools}
\usepackage[normalem]{ulem}
\usepackage{simplewick}

\def\beq{\begin{equation}}
\def\eeq{\end{equation}}
\def\beqa{\begin{eqnarray}}
\def\eeqa{\end{eqnarray}}

\begin{document}

\title{\Large Lensing with generalized symmetrons}

\author{Christian K\"{a}ding}
\email{christian.kaeding@tuwien.ac.at}
\affiliation{Technische Universit\"at Wien, Atominstitut, Stadionallee 2, 1020 Vienna, Austria}

\begin{abstract}
Generalized symmetrons are models that have qualitatively similar features to the archetypal symmetron, but have barely been studied. In this article, we investigate for what parameter values the fifth forces induced by disformally coupling generalized symmetrons can provide an explanation for the difference between baryonic and lens masses of galaxies. While it is known that the standard symmetron struggles with providing an alternative source for the lensing otherwise attributed to particle dark matter, we show that some generalized symmetron models are more suitable for complying with existing constraints on disformal couplings. This motivates future studies of these only little explored models.

\end{abstract}


\maketitle



\section{Introduction}

Some of modern physics' most prominent open problems can be found in cosmology, i.e.\,\,the questions after the natures of dark energy (DE) and dark matter (DM). In order to tackle these problems, a zoo of modifications of general relativity have been considered. Amongst those, scalar-tensor theories \cite{Fujii2003} are some of the most studied ones. An overview of models that address the problems of DE and DM can be found in Refs.\,\,\cite{Clifton2011,Joyce2014}. 
\\
Some of  the scalar fields appearing in scalar-tensor theories are expected to cause a fifth force of Nature, which, however, is in tension with Solar System-based experiments \cite{Dickey1994,Adelberger2003,Kapner2007}. This led to some of these models being already ruled out \cite{Ishak2018}. Though, so-called screened scalar fields, see Refs.\,\,\cite{BurrageSak,Brax:2021wcv} for reviews, have ways of circumventing Solar System constraints by screening their fifth forces in environments of higher mass densities, such that the forces are effectively rendered to be feeble. The most well-known screened scalar field models include the chameleon \cite{Khoury2003,Khoury20032}, the symmetron \cite{Dehnen1992, Gessner1992, Damour1994, Pietroni2005, Olive2008, Brax2010,Hinterbichler2010,Hinterbichler2011}, the environment dependent dilaton \cite{Damour1994,Gasperini:2001pc,Damour:2002nv,Damour:2002mi,Brax2010,Brax:2011ja,Brax2022}, and the galileon \cite{Dvali2000,Nicolis2008,Ali2012}. In recent years these models have been (proposed to be) tested in various experiments, see e.g. Refs.\,\,\cite{Burrage:2016bwy,BurrageSak,Pokotilovski:2012xuk,Pokotilovski:2013tma,Burrage:2014oza,Hamilton:2015zga,Lemmel:2015kwa,Burrage:2015lya,Elder:2016yxm,Ivanov:2016rfs,Burrage:2016rkv,Jaffe:2016fsh,Brax:2017hna,Sabulsky:2018jma,Brax:2018iyo,Cronenberg:2018qxf,Hartley2019,Pitschmann:2020ejb}, studied as quantum fields \cite{Brax2018quantch,Burrage2018,Burrage2019,Kading2019}, and proposed for investigations in analogue gravity simulations \cite{Hartley2018}. 
\\
Screened scalar field models have a rich phenomenology, which allows them to serve as promising candidate theories for DM or DE. For example, in Ref.\,\,\cite{Burrage2016_2} it was shown that the symmetron fifth force can explain the stability and rotation curves of disk galaxies, while Ref.\,\,\cite{OHare:2018ayv} demonstrated how the same force can lead to the observed motion of stars perpendicular to the plane of the Milky Way disk. This made the symmetron a promising alternative to particle DM. Though, in Refs.\,\,\cite{Burrage2018Sym} and \cite{Kading2019} it was shown that, using the same parameter values as in Ref.\,\,\cite{Burrage2016_2}, the difference between the baryonic masses of galaxies and the masses required for causing the observed (gravitational) lensing cannot be explained by a symmetron fifth force alone, but potentially by adding another scalar field or considering a hybrid model between modified gravity and particle DM. 
\\
In the present article we revisit the idea of a symmetron fifth force being an alternative to particle DM. However, instead of only considering the archetypal symmetron, we discuss so-called generalized symmetrons, which were discovered using tomographic methods \cite{Brax2011,Brax2012}. This class of models comprises the standard symmetron, but also an, in principle, infinite number of qualitatively similar generalizations. Generalized symmetrons have barely been studied in the literature \cite{BurrageSak} and their parameter spaces are still unconstrained. As we show in the present article, some generalized symmetron models can actually explain the observed lensing otherwise attributed to particle DM while complying with existing constraints \cite{Brax:2014vva,Brax:2015hma} on the necessary disformal coupling \cite{Bekenstein:1992pj}. Since the parameter spaces of those models have yet not been constrained by experiments, they offer significantly more freedom to discuss them as alternatives to particle DM than many other screened scalar field models including the standard symmetron do. This motivates future studies of these yet only little explored models.
\\
The article is structured as follows: In Sec.\,\,\ref{sec:Sym} we review the generalized symmetron models, while in Sec.\,\,\ref{sec:Lensing}, based on the discussion in Ref.\,\,\cite{Kading2019}, we derive the deviation of a galaxy's lensing from general relativity's predictions due the presence of a generalized symmetron fifth force. Subsequently, in Sec.\,\,\ref{sec:Param}, we study which parts of the generalized symmetron parameter spaces are suitable for explaining the difference between baryonic and lens masses of galaxies. Finally, we draw our conclusions in Sec.\,\,\ref{sec:Conclusion}. 


\section{Generalized symmetrons}
\label{sec:Sym}

\begin{figure}[htbp]
\begin{center}
\includegraphics[scale=0.3]{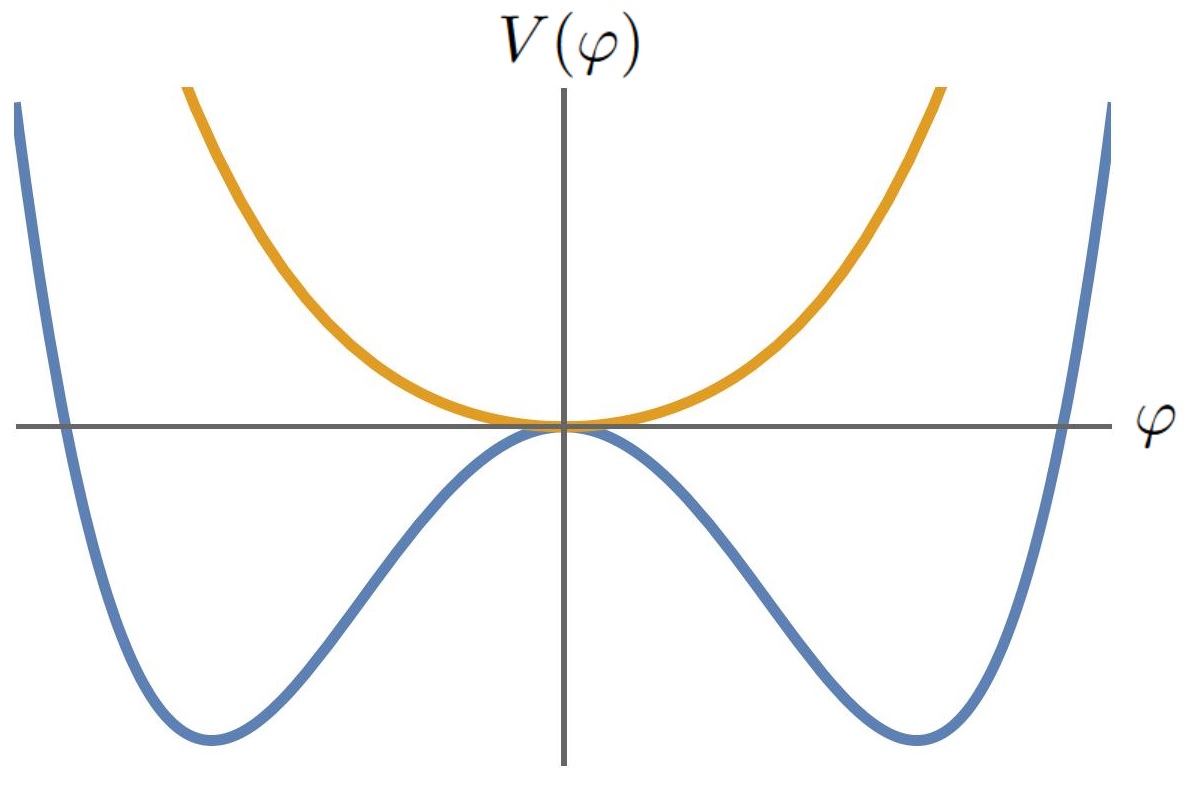}
\caption{Potential $V(\varphi)$ of the (generalized) symmetron: if the density $\rho$ is sufficiently small, then the potential takes on the shape of the blue curve and the field has a non-vanishing vev. However, if $\rho>\mu^2\mathcal{M}^2$ (orange curve), then the potential can only have a minimum at $\varphi=0$ , which results in the fifth force being screened.}
\label{fig:Profil}
\end{center}
\end{figure}

The standard symmetron model is a scalar-tensor modification of gravity and introduces a new type of scalar field - the symmetron. It was first mentioned in Refs.\,\,\cite{Dehnen1992, Gessner1992, Damour1994, Pietroni2005, Olive2008, Brax2010} and then introduced with its current name in Refs.\,\,\cite{Hinterbichler2010,Hinterbichler2011}. Besides being a potential DM candidate, the symmetron also motivated a new inflationary scenario \cite{Dong:2013swa} and was proposed as a solution to the $H_0$-tension \cite{Solomon:2022qqf}.
\\
The symmetron $\varphi$ is described by the action
\begin{eqnarray}
S &=& \int d^4x \sqrt{-g}\left( \frac{M_P^2}{2}R - \frac{1}{2}(\partial\varphi)^2 - V(\varphi) \right) \,\,\,, 
\end{eqnarray}
where $M_P$ is the reduced Planck mass, and the effective symmetron potential is given by
\begin{eqnarray}\label{eq:Standard}
    V(\varphi) &=& \frac{1}{2}\left( \frac{\rho}{\mathcal{M}^2} -\mu^2 \right)\varphi^2 + \frac{\lambda}{4}\varphi^4\,\,\,.
\end{eqnarray}
Here, $\rho = -T^\mu_{\,\,\,\mu}$ is the density for dust-like matter, $\mathcal{M}$ is a constant with dimension of a mass that parametrizes the coupling to matter, $\mu$ is a tachyonic mass, and $\lambda$ is the dimensionless self-coupling constant of the symmetron. The universal coupling to the trace of the matter energy-momentum tensor $T^\mu_{\,\,\,\mu}$ arises since the symmetron couples conformally to the metric tensor via a factor
\begin{eqnarray}
    A(\varphi) &=& 1 + \frac{\varphi^2}{\mathcal{M}^2} + \mathcal{O}\left(  \frac{\varphi^4}{\mathcal{M}^4}\right)\,\,\,,\,\,\,\varphi\ll\mathcal{M}\,\,\,.
\end{eqnarray}
This factor is used in order to translate from one conformal frame \cite{Fujii2003} to another, e.g.\,\,from the Einstein frame given in terms of the metric $g$ to the Jordan frame given in terms of $\tilde{g}$ via $\tilde{g}_{\mu\nu} = A(\varphi)g_{\mu\nu}$.
\\
Due to its coupling to matter, an additional force of Nature, a so-called fifth force, is expected to be mediated by the symmetron. Though, in order to avoid Solar System constraints on fifth forces \cite{Dickey1994,Adelberger2003,Kapner2007}, the symmetron is equipped with a screening mechanism, which renders the fifth force to be feeble in regions where the trace of the energy-momentum tensor, or, in case of dust, the density $\rho$, is sufficiently large. This screening is achieved by the behavior of the symmetron's non-linear effective potential given in Eq.\,\,(\ref{eq:Standard}). As long as $\mu^2\mathcal{M}^2 > \rho$, this potential has minima at
\begin{eqnarray}
    \varphi_0 &=& \pm \sqrt{\frac{1}{\lambda}\left(\mu^2 -  \frac{\rho}{\mathcal{M}^2}\right)}\,\,\,.
\end{eqnarray}
However, if this condition is not fulfilled, the vacuum expectation value (vev) of the symmetron can only be $\varphi_0 =0$. Both possible scenarios are depicted in Fig.\,\ref{fig:Profil}. Separating the symmetron into its vev and a small fluctuation $\delta\varphi$, such that $\varphi = \varphi_0 + \delta\varphi$, it can be shown that the symmetron fifth force at leading order goes like $F_\varphi \sim \varphi_0 \nabla \delta\varphi$. Consequently, in situations where $\mu^2\mathcal{M}^2 < \rho$, the symmetron-induced fifth force is screened since its leading term vanishes, leaving only small corrections of higher order in $\delta\varphi$.
Besides this standard symmetron model, a radiatively-stable symmetron has been developed \cite{Burrage2016}, and generalizations were discovered by using tomographic methods \cite{Brax2011,Brax2012}. The latter are the main subject of this article and discussed in what follows. 
Generalized symmetrons couple to the metric tensor via
\begin{eqnarray}\label{eq:ConfParam}
    A(\varphi) &=& 1 + \frac{\varphi^{2\alpha}}{\mathcal{M}^{2\alpha}} + \mathcal{O}\left(  \frac{\varphi^{4\alpha}}{\mathcal{M}^{4\alpha}}\right)\,\,\,,\,\,\,\varphi\ll\mathcal{M}\,\,\,,
\end{eqnarray}
and have potentials \cite{BurrageSak}
\begin{eqnarray}\label{eq:General}
V(\varphi) &=& \left( \frac{\rho}{\mathcal{M}^{2\alpha}} - \mu^{4-2\alpha} \right) \varphi^{2\alpha} + \frac{\varphi^{2\beta}}{\Lambda^{2\beta-4}}\,\,\,,
\end{eqnarray}
where $\Lambda$ describes a symmetron self-coupling constant which generally has the dimension of a mass, but is replaced $1/\Lambda^{2\beta-4} \to \lambda/4$ for $\beta =2$. Note that, for $\alpha =2$, also $\mu$ must be replaced by a dimensionless constant. The numbers $\alpha, \beta \in \mathbb{Z}^+$ label each model and are, in principle, arbitrary, but must adhere to $\beta > \alpha$. Choosing the smallest possible pair $(\alpha,\beta) = (1,2)$ recovers the standard symmetron model as given in Eq.\,\,(\ref{eq:Standard}). Each choice for $(\alpha,\beta)$ leads to a model that has similar qualitative features to the standard symmetron. This means, the potential in Eq.\,\,(\ref{eq:General}) is also represented by Fig.\,\,\ref{fig:Profil}, a generalized symmetron fifth force is screened in regimes where $\rho > \mu^{4-2\alpha}\mathcal{M}^{2\alpha}$ since there $\varphi_0 = 0$, and in unscreened regimes the field's vev takes on the form  
\begin{eqnarray}\label{eq:vev}
\varphi_0 &=& \pm\sqrt[2(\beta-\alpha)]{\frac{\alpha}{\beta}\Lambda^{2\beta-4}\left( \mu^{4-2\alpha} - \frac{\rho}{\mathcal{M}^{2\alpha}}  \right)}\,\,\,.
\end{eqnarray}
In an unscreened situation a generalized symmetron has a mass 
\begin{eqnarray}\label{eq:unscrmass}
    m^2 &=& 4\beta(\beta-\alpha)\frac{\varphi_0^{2(\beta-1)}}{\Lambda^{2\beta-4}}\,\,\,,
\end{eqnarray}
while in case of total screening, this becomes 
\begin{eqnarray}
    m^2 &=& 
    \begin{cases}
    \frac{\rho}{\mathcal{M}^2}-\mu^2 &,\,\,\, \alpha =1
    \\
    0 &,\,\,\, \alpha >1
    \end{cases}\,\,\,.
\end{eqnarray}


\section{Lensing}
\label{sec:Lensing}

\begin{figure}[htbp]
\begin{center}
\includegraphics[scale=0.60]{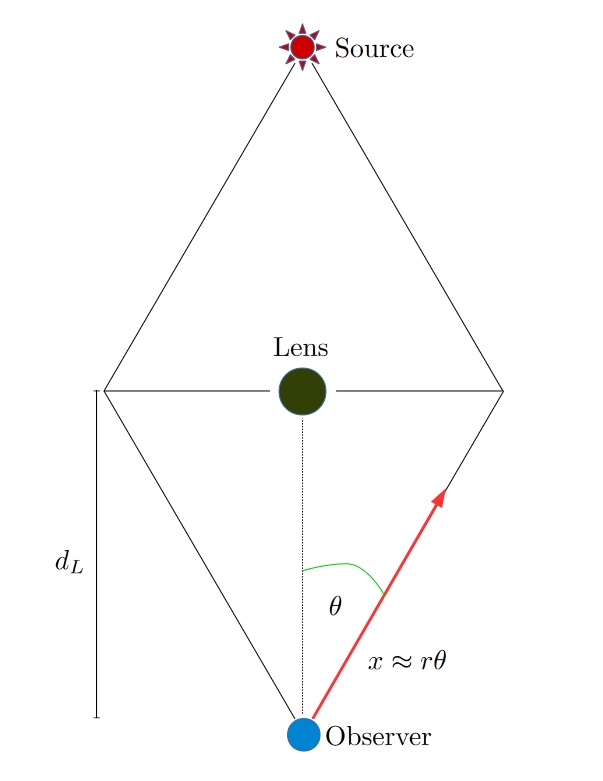}
\caption{Schematics of gravitational lensing in two dimensions with thin-lens approximation: The source sends out light which is distorted by the lens and then received by an observer. $d_L$ denotes the distance between the observer and the lens plane in which the lens is situated. The radial coordinate $r$
equals $0$ at the position of the observer and is orthogonal to the lens plane. $\theta$ denotes the angle between $r$ and the light ray reaching the observer (here
it represents either $\theta^1$ or $\theta^2$). The coordinate $x$ is approximated by $r\theta$ since
$\theta$ is assumed to be very small. \cite{Kading2019}}
\label{fig:Lensing}
\end{center}
\end{figure}

Gravitational lensing became the first experimentally confirmed novel effect predicted by general relativity when in 1919 the bending of light in the Sun's gravitational field was observed \cite{Dyson:1920cwa}. Today gravitational lensing serves as a valuable tool for indirect observations of DM \cite{Massey:2010hh}. If a symmetron fifth force is supposed to act as an alternative to particle DM, as suggested by the findings in Refs.\,\,\cite{Burrage2016_2,OHare:2018ayv}, then it must be able to explain the observed roughly $1:5$ ratio between baryonic mass and DM in galaxies \cite{Weinberg_2008}, i.e.\,\,contribute to the lensing about $5$ times as much as the Newtonian potential $\Phi$ of a galaxy. In Refs.\,\,\cite{Burrage2018Sym} and \cite{Kading2019} it was shown that the standard or $(1,2)$-symmetron cannot provide such an explanation for the parameter values required by Ref.\,\,\cite{Burrage2016_2}. Though, other models of the class of generalized symmetrons might still be able to serve as sensible alternatives to particle DM since they show the same qualitative features as the $(1,2)$-symmetron, but  are yet much less constrained.  
\\
In order to study the effect of a generalized symmetron on lensing, we now derive a measure that enables us to compare the contribution from the fifth force with the one from Newtonian gravity. For this, we follow the discussion in Ref.\,\,\cite{Kading2019}, which in turn is based on elaborations made in Ref.\,\,\cite{Amendola:2015ksp}:
\\
We consider a perturbed Friedmann-Lema\^itre-Robertson-Walker (FLRW) background 
\begin{eqnarray}
ds^2 &=& a^2(\tau)[-(1+2\Psi)d\tau^2 + (1+2\Phi)\delta_{ij}dx^idx^j]
\end{eqnarray}
with conformal time $d\tau = dt/a(t)$ and Newtonian potentials $\Phi, \Psi \ll 1$, which fulfill the no-slip condition $\Phi = - \Psi$. Furthermore, we assume that the thickness of the lens (e.g.\,\,a galaxy) is much smaller than the distance $d_L$ between lens and observer (on Earth), and the deflection angles are very small. This assumption justifies a thin-lens approximation, which effectively considers the lens to have no thickness and lie within a two-dimensional plane. This so-called lens plane is perpendicular to the line between light source and observer. Using this approximation, we introduce the coordinate system
$
\{ \tau, r, x^1 \approx r\theta^1, x^2 \approx r\theta^2 \}
$
with radial coordinate $r$ and deflection angles $\theta^{1,2}$. See Fig.\,\,\ref{fig:Lensing} for a depiction of the setup.
\\
We introduce the photon momentum $k^\mu := dx^\mu/d\kappa$ with some affine parameter $\kappa$, which gives the geodesic equation
\begin{eqnarray}\label{eq:geodesic}
\frac{dk^\mu}{d\kappa} + \Gamma^\mu_{\nu\sigma}k^\nu k^\sigma &=& 0\,\,\,.
\end{eqnarray}
The photon momentum can be separated into a background vector $\hat{k}^\mu$ and a small perturbation $\delta k^{\mu}$ due to the presence of a lens mass, such that $k^\mu = \hat{k}^\mu + \delta k^{\mu}$. Without the lens mass, the photon trajectory is not subject to deflection, which means, for the background momentum $\hat{k}^{x^{1,2}}=0$ must hold. Consequently, from the on-shell condition $k^\mu k_\mu=0$ follows $\hat{k}^r = \hat{k}^0$.
\\
Taking all this into account, and considering only terms up to first order in the Newtonian potentials and the momentum perturbations, for the purely Newtonian case without any fifth forces, the geodesic Eq.\,\,(\ref{eq:geodesic}) can be used to derive the lensing force law
\begin{eqnarray}\label{eq:FLNEwton}
\frac{d^2x^i}{dr^2} &=& (\Phi-\Psi),_{x^i}\,\,\,,\,\,\,i\in\{1,2\}\,\,\,.
\end{eqnarray}
A similar force law can be derived when also including generalized symmetrons. Though, it must be stressed that solely a conformal coupling of the scalar field $\varphi$ to the metric tensor is not sufficient to have any effect on lensing. This was, for example, explicitly shown in Ref.\,\,\cite{Kading2019}, but can also simply be concluded from the fact that the energy-momentum tensor of a massless particle like the photon has a vanishing trace. Therefore, the scalar $\varphi$ must not only couple conformally, but also disformally via 
\begin{eqnarray}
\bar{g}_{\mu\nu} & = & A(\varphi)g_{\mu\nu} +B\varphi,_{\mu}\varphi,_{\nu}\,\,\,,
\\
\bar{g}^{\mu\nu} & = & A^{-1}(\varphi)\left( g^{\mu\nu}-\frac{B}{C(\varphi)} \varphi,^{\mu}\varphi,^{\nu} \right) 
\,\,\,,
\end{eqnarray}
where $A(\varphi)$ is the conformal factor from Eq.\,\,(\ref{eq:ConfParam}), $B$ is the disformal coupling parameter, which for simplicity is assumed to be a constant, and $C(\varphi):=A(\varphi)+B(\partial\varphi)^2$. $\bar{g}$ denotes the metric in the Jordan frame and $g$ the metric in the Einstein frame \cite{Fujii2003}.
\\
For the Jordan frame Christoffel symbols we find
\begin{eqnarray}\label{eq:Christoffel}
\bar{\Gamma}^\mu_{\nu\sigma} 
& = & \Gamma^\mu_{\nu\sigma} + A^{-1}\left[A,_{(\nu}g^{\mu}_{\sigma)} -\frac{1}{2} A,^{\mu}g_{\nu\sigma}\right]
\nonumber
\\
&&
+ \frac{B}{C} \varphi,^\mu\left[\varphi,_{\nu\sigma} - \varphi,_\rho \Gamma^\rho_{\nu\sigma} 
+\frac{A^{-1}}{2}\varphi,^\rho A,_\rho g_{\nu\sigma}-A^{-1} A,_{(\nu} \varphi,_{\sigma)}\right]\,\,\,,
\end{eqnarray}
where the second term in the first line represents a contribution which is arising purely from the conformal coupling and therefore does not affect the lensing. Substituting Eq.\,\,(\ref{eq:ConfParam}) into Eq.\,\,(\ref{eq:Christoffel}), following the same procedure as for the derivation of Eq.\,\,(\ref{eq:FLNEwton}), and assuming the lens object to be static, then the force law 
\begin{eqnarray}\label{eq:FLSym}
\frac{d^2x^i}{dr^2} &=& (\Phi - \Psi),_{x^i} 
- \frac{B}{C} \varphi,^{x^i}\left[ \bigg(\varphi,_{yz} - \frac{2\alpha}{\mathcal{M}^{2\alpha}}\varphi^{2\alpha -1}\varphi,_y \varphi,_z\bigg)\frac{dx^y}{dr} \frac{dx^z}{dr}\right.
 \nonumber
\\
&&
\,\,\,\,\,\,\,\,\,\,\,\,\,\,\,\,\,\,\,\,\,\,\,\,\,\,\,\,\,\,\,\,\,\,\,\,\,\,\,\,\,\,\,\,\,\,\,\,\,\,\,\,
 \left.  - 2\mathcal{H} \varphi,_y \frac{dx^y}{dr} + a^2 \varphi,_y  (\Phi - \Psi),^y -2\varphi,_r \Phi,_r  \right]\,\,\,, 
\end{eqnarray}
where $y,z \in\{ r, x^1, x^2 \}$, and $\mathcal{H}$ denotes the conformal Hubble parameter, can be obtained. Comparing this force law with the one in Eq.\,\,(\ref{eq:FLNEwton}), we find that the term proportional to $B$ in Eq.\,\,(\ref{eq:FLSym}) corresponds to the contribution of the generalized symmetron fifth force to the lensing effect.
\\
Next, we assume $a(\text{today}) = 1$, use the no-slip condition for the Newtonian potentials, and consider the force law only in the lens plane $r=d_L$, where we expect the lensing to be maximal and $dx/dr$ to vanish:
\begin{eqnarray}\label{eq:LPForceL}
\left. \frac{d^2x^i}{dr^2} \right|_{d_L} &=& \bigg[ 2\Phi,_{x^i} 
\left. 
- \frac{B}{C}\varphi,^{x^i} \bigg( \varphi,_{rr} - \frac{2\alpha}{\mathcal{M}^{2\alpha}} \varphi^{2\alpha-1}(\varphi,_r)^2 - 2\mathcal{H}\varphi,_r +  2\varphi,_{x^j} \Phi,^{x^j}\bigg) \bigg]\right|_{d_L}\,\,\,.
\end{eqnarray}
In order to get a rough numerical estimate of the generalized symmetron contribution, we assume its source, i.e.\,\,a galaxy acting as lens, to be a disk lying in the lens plane with homogeneous, constant mass density and radius $R$, leading to a total galaxy mass $M$. Since this approximated lens has a spherical symmetry, we can restrict our investigation to the case $\theta^2 =0$, and consequently only work with $\theta^1 =:\theta$. Applying Newton's shell theorem and considering that the radial coordinate within the lens plane originating from the disk's center can be expressed as $r\theta$ in the observer's coordinates, the Newtonian potential becomes
\begin{eqnarray}\label{eq:NewPot}
    \Phi &=& -\frac{GM}{r\theta}\,\,\,.
\end{eqnarray}
Some references, including Refs.\,\,\cite{Kading2019,Brax2012}, suggest an approximation for the symmetron field profile outside a homogeneous, spherically symmetric source of radius $R$, which is valid for a symmetron mass $m_\text{out}$ outside the source fulfilling $r\theta < m_\text{out}^{-1}$. However, in what follows, this approximation will not always be good. Therefore, we use the symmetron profile \cite{Burrage:2016rkv}
\begin{eqnarray}\label{eq:SymProfile}
    \varphi &=& v - \frac{D}{r\theta} e^{-m_\text{out}r\theta}
\end{eqnarray}
with
\begin{eqnarray}\label{eq:D}
    D &:=& (v-w) R e^{m_\text{out} R} \frac{R m_\text{in} - \mathcal{T}}{Rm_\text{in} + R m_\text{out} \mathcal{T} }\,\,\,,
\end{eqnarray}
where $m_\text{in}$ is the field's mass within the source,  $v:=\varphi_{0,\text{out}}$ and $w := \varphi_{0,\text{in}}$ are the vev in the environment surrounding the source and within the source, respectively, and $\mathcal{T} := \tanh(m_\text{in}R)$. The solution in Eq.\,\,(\ref{eq:SymProfile}) requires that we only consider terms up to first order in $\delta\varphi$, and, at this order, is even valid for any generalized symmetron model. 
\\
Substituting Eqs.\,\,(\ref{eq:NewPot}) and (\ref{eq:SymProfile}) into Eq.\,\,(\ref{eq:LPForceL}) leads us to
\begin{eqnarray}
\left. \frac{d^2x}{dr^2} \right|_{d_L} &=& \frac{2GM}{d^2_L\theta^2}  
\left[1+ F\right]
\end{eqnarray}
with
\begin{eqnarray}\label{eq:FuncF}
F &:=& 
\frac{BD^2}{Cd_L^4\theta^2 } e^{-2m_\text{out} d_L \theta} (1+m_\text{out}d_L\theta)
\bigg\{\left( 1+ \frac{d_L\theta}{2GM} \right) \bigg[1 + (1+m_\text{out}d_L\theta)^2
\nonumber
\\
&&
+ \frac{2\alpha D}{d_L\theta \mathcal{M}^{2\alpha}}
\left(v - \frac{D}{d_L\theta} e^{-m_\text{out}d_L\theta}\right)^{2\alpha-1}e^{-m_\text{out}d_L\theta} (1+m_\text{out}d_L\theta)^2
+2 \mathcal{H}d_L(1+m_\text{out}d_L\theta)\bigg]
\nonumber
\\
&&
\,\,\,\,\,\,\,\,\,\,\,\,\,\,\,\,\,\,\,\,\,\,\,\,\,\,\,\,\,\,\,\,\,\,\,\,\,\,\,\,\,\,\,\,\,\,\,\,\,\,\,\,\,\,\,\,\,\,\,\,\,\,\,\,\,\,\,\,\,\,\,\,\,\,\,\,\,\,\,\,\,\,\,\,
- \frac{1+m_\text{out}d_L\theta}{\theta^2}\bigg\}
\end{eqnarray}
being a term that describes the contribution of the generalized symmetron fifth force to lensing in comparison to the one originating from Newtonian gravity. Furthermore, in the lens plane we have:
\begin{eqnarray}
C|_{d_L} 
&=& 1+ \frac{1}{\mathcal{M}^{2\alpha}}\left(v - \frac{D}{d_L\theta} e^{-m_\text{out}d_L\theta}\right)^{2\alpha}
+(1-2\Phi|_{d_L}) \frac{BD^2}{d_L^4\theta^4}
e^{-2m_\text{out}d_L\theta}
(1+m_\text{out}d_L\theta)^2(1+\theta^{2}) \,\,\,.
\nonumber
\\
\end{eqnarray}
In Eq.\,\,(\ref{eq:FuncF}) we see that $F \sim D^2$, such that from Eq.\,\,(\ref{eq:D}) we can find $F \sim (v-w)^2$. This result implies that if $v$ and $w$ are very similar, and have the same sign, for example in some situations where the field is unscreened both in- and outside the source, the contribution of the generalized fifth force to lensing can be very small. However, it is not necessarily true that both vev need to have the same sign. Since we are interested in checking whether it is at all possible to explain the observed lensing by a generalized symmetron fifth force, we consider the best case scenario, in which $v = +|v|$ and $w= -|w|$.


\section{Model parameters}
\label{sec:Param}

We now want to consider different generalized symmetron models and see for what points in their parameter spaces the expression in Eq.\,\,(\ref{eq:FuncF}) gives $F \approx 5$, such that the contribution of a fifth force can be interpreted as an alternative to particle DM, at least when it comes to lensing.
\\
For this, we consider the same galaxy parameters as in Ref.\,\,\cite{Kading2019}, i.e.\,\,we look at a Milky Way-like galaxy with $M = 6 \times 10^{11} M_\odot \approx 6.67 \times 10^{77} \,\text{eV}$ and a scale length $R = 5\, \text{kpc} \approx 2.69 \times 10^{26} \,\text{eV}^{-1}$ \cite{Fathi_2010}. Ref.\,\,\cite{DES:2017gwu} reports lensing by galaxies at redshift $z=1$ under angles of about $1\,\text{arcmin}$. Therefore, we choose $\theta = \frac{\pi}{10800}$ and $d_L \approx 6.60\times 10^{32} \,\text{eV}^{-1}$, where we obtained the latter from $d_L \approx z d_H$ \cite{Hogg:1999ad} using the Hubble length $d_H$. For the density around the galaxy we assume $\rho_\text{out} \approx 2.59\times10^{-11}\,\text{eV}^{4}$ \cite{Prat:2021xlz}. In addition, we use $G \approx 6.71 \times 10^{-57} \,\text{eV}^{-2}$ and $\mathcal{H} \approx 1.51 \times 10^{-33} \,\text{eV}$.
\\
We start the discussion by again looking at the $(1,2)$-symmetron before we show that moving to larger values of $(\alpha,\beta)$ is beneficial for complying with the existing constraint $B < 5.6 \times 10^{-48} \,\text{eV}^{-4}$ \cite{Brax:2014vva,Brax:2015hma} on the disformal coupling parameter.

\subsection{$(1,2)$-symmetron}

\begin{figure}[htbp]
\begin{center}
\includegraphics[scale=0.80]{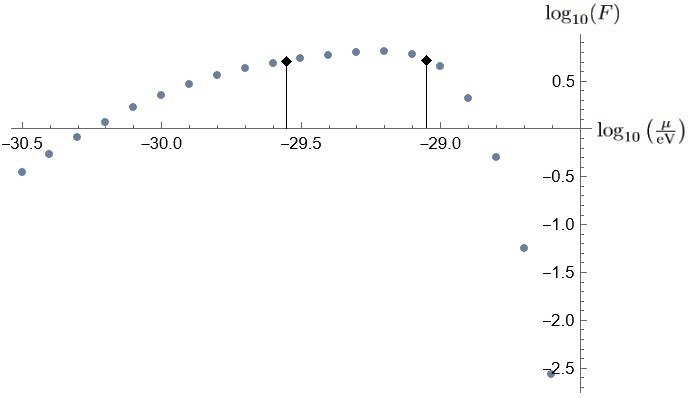}
\caption{Behavior of the function $F(\mu)$ for the $(1,2)$-symmetron with parameters $\lambda = 10^{-174.6}$, $\mathcal{M} = 10^{58.7}\,\text{eV}$, and $B=10^{-49}\,\text{eV}^{-4}$; the black rhombuses with vertical lines depict the two points where $F =5$.    }
\label{fig:1_2_mu}
\end{center}
\end{figure}

Ref.\,\,\cite{Burrage2016_2} suggested that the $(1,2)$-symmetron could explain the stability and rotation curves of disk galaxies, and therefore be a possible candidate for an alternative to particle DM. For this to work, the symmetron parameters were chosen to be  $\mathcal{M} = M_P/10$, $v = \mathcal{M}/150$, and $\mu = 3 \times10^{-30} \,\text{eV}$. In Refs.\,\,\cite{Burrage2018Sym} and \cite{Kading2019} it was shown that those parameter choices require a disformal coupling parameter $B$ much larger than permitted by the constraints given in Refs.\,\,\cite{Brax:2014vva,Brax:2015hma}. This result also held for realistic variations of the galaxy parameters. 
\\
We now study for what values of $\mathcal{M}$ and the self-coupling constant $\lambda$ the $(1,2)$-symmetron can actually comply with the constraints on disformal couplings. For the disformal coupling parameter we choose $B=10^{-49}\,\text{eV}^{-4}$ in order to consider a value that is not yet excluded by experiments, and for $\mu$ we initially take the same value as in Ref.\,\,\cite{Burrage2016_2}. We find that $\lambda \approx 10^{-174.6}$ is the largest and $\mathcal{M} \approx 10^{58.7}\,\text{eV}$ the smallest possible value in order to find $F\approx 5$, while simultaneously fulfilling the perturbative condition $\varphi \ll \mathcal{M}$. Even though there are not necessarily any restrictions on the permitted values for the coupling constants, it is certainly peculiar to have a theory with such a small self-coupling and, more strikingly, a mass scale that is more than $30$ orders of magnitude above the Planck mass. 
\\
Note that for these parameter values the field is not screened within the galaxy. Furthermore, since in this parameter regime $1/m_\text{in} > R$, i.e.\,\,the Compton wavelength of the field is larger than the galaxy scale, the field adapts to the size of the galaxy and consequently rather takes on the mass $m_\text{in} \approx q/R$ and the vev
\begin{eqnarray}
    w &\approx& \pm
    \sqrt[\beta -1]{\frac{q \Lambda^{\beta-2}}{2\sqrt{\beta(\beta-\alpha)}R}}
    \,\,\, ,
\end{eqnarray}
where $q$ is a fudge factor that would have to be computed numerically taking into account more detailed properties of a galaxy, but is assumed to be of order $1$ (compare with the fudge factor, e.g., in Ref.\,\,\cite{Brax:2016wjk}).
\\
Considering smaller values of the disformal coupling parameter only worsens the situation, i.e.\,\,requires smaller values of $\lambda$ and larger $\mathcal{M}$.  As was also observed in Ref.\,\,\cite{Kading2019}, changing the galaxy parameters to other realistic values does not significantly improve the situation. 
\\
What remains to be checked is how the result is affected by a change in the tachyonic mass $\mu$. As it turns out, and can be seen in Fig.\,\,\ref{fig:1_2_mu}, the value for $\mu$ used in Ref.\,\,\cite{Burrage2016_2} is sitting within a small part of the parameter space that allows for $F \approx 5$. 
\\
Looking at Eqs.\,\,(\ref{eq:SymProfile}) and (\ref{eq:D}), and remembering that $F \sim D^2$, we see that significantly increasing or decreasing $\mu$ beyond this small section of the parameter space, while keeping all other parameters at the same values, leads to $F$ getting rapidly smaller since the combined exponential function in Eqs.\,\,(\ref{eq:SymProfile}) and (\ref{eq:D}) becomes smaller due to the enlarged symmetron mass or the vev reduces, respectively. In order to counteract the declining $F$, such that $F \approx 5$ can be recovered, we have to consider even smaller values of $\lambda$ and consequently larger values of $\mathcal{M}$. This is the exact opposite of what we had hoped to achieve, which is why varying $\mu$ can be excluded as a possibility for improving the case for the $(1,2)$-symmetron.

\subsection{$(1,\beta)$-symmetrons with $\beta \geq 3$}

\begin{figure}[htbp]
\begin{center}
\includegraphics[scale=0.80]{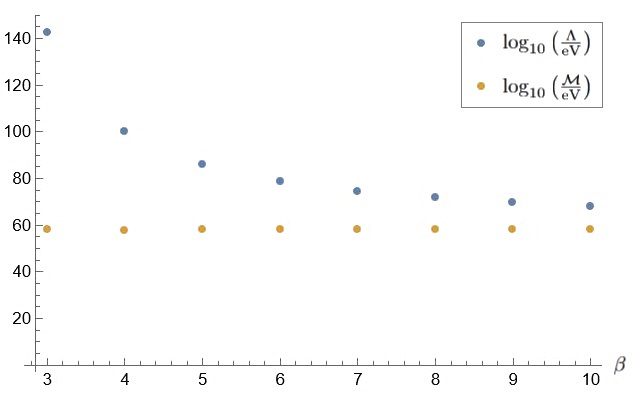}
\caption{Values of $\Lambda$ and $\mathcal{M}$ that give $F = 5$ for $(1,\beta \geq 3)$-symmetrons with $\mu = 3 \times10^{-30} \,\text{eV}$ up to $\beta =10$}
\label{fig:1_3_beta}
\end{center}
\end{figure}

Now we move to the first set of non-standard symmetrons and consider models with $(1,\beta)$ for $\beta \geq 3$. For those, Eq.\,\,(\ref{eq:FuncF}) has still the same form as for the $(1,2)$-symmetron, but the vev and the unscreened masses are different according to Eqs.\,\,(\ref{eq:vev}) and (\ref{eq:unscrmass}). Furthermore, instead of using a dimensionless constant $\lambda$, we now have to work with $\Lambda$ as another mass scale besides $\mathcal{M}$. Again choosing $\mu = 3 \times10^{-30} \,\text{eV}$, we find the results presented in Fig.\,\,\ref{fig:1_3_beta} for $3 \leq \beta \leq 10$.
\\
We observe that going from $\beta =2$ to $\beta = 3$ requires a slightly lower $\mathcal{M}$ for $F \approx 5$ and $\varphi/\mathcal{M}\ll 1$, but the $\Lambda$ of the $(1,3)$-symmetron must be more than $100$ orders of magnitude above the Planck mass. This renders this theory to be unrealistic as an alternative to particle DM.
\\
At $\beta =4$ there is another slight decrease in $\mathcal{M}$, while from $\beta =5$ on this mass scale increases again until it settles around $\mathcal{M} \approx 10^{58.3}\,\text{eV}$ even for very large $\beta$. This means, compared to the standard symmetron, only changing $\beta$ barely improves the value of the matter coupling mass scale.
\\
The self-coupling constant $\Lambda$ is strictly monotonically decreasing and approaches $\Lambda \approx 10^{57.2}\,\text{eV}$ for large $\beta$. This means, the two mass scales in $(1,\beta \geq 3)$-symmetron theories are getting closer to each other with increasing $\beta$ until $\Lambda$ becomes smaller than $\mathcal{M}$ and both finally differ by approximately $1$ order of magnitude.
\\
Even though there are parts of their parameter spaces for which generalized symmetrons with $\alpha =1$ can explain the lensing otherwise attributed to particle DM, they require coupling constants which are at least $30$ order of magnitude above the Planck mass. 

\subsection{$(\alpha,\beta)$-symmetrons with $\alpha \geq 2$}

Finally, we also consider generalized symmetron models with $\alpha \geq 2$. Now we encounter a couple of differences to the previously considered models: Eq.\,\,(\ref{eq:FuncF}) changes with every possible value of $\alpha$, and $\mu$ is either dimensionless (for $\alpha =2$) or appears in Eq.\,\,(\ref{eq:General}) with negative order (for $\alpha \geq 3$).
\\
Generally, it can be said that studying these models is numerically more intricate than it was for the ones we considered before. This is due to the fact that, within the considered astrophysical setup, $\alpha \geq 2$-symmetrons can lead to extremely small terms that have to be compensated by extremely large terms in order to find any results of order $1$. For example, Eq.\,\,(\ref{eq:FuncF}) can be separated into two terms, one outside ($F_o$) and one inside ($F_i$) the curly brackets. $F_o$ tends to become very small since it is suppressed by at least $B/d_L^3\theta \sim 10^{-144}\,\text{eV}^{-1}$ (for $B=10^{-49}\,\text{eV}^{-4}$). While generalized symmetrons with only $\alpha =1$ seem to not struggle with compensating such smallness by leading to a sufficiently large $F_i$, the models discussed here are more problematic since they lead to even more extreme values for $F_o$ and $F_i$.
\\
Beginning with the $(2,3)$-symmetron model, we can find no point in the parameter space that allows for $F\approx 5$. The closest value we find is around $F\sim 10^{-139}$. This is caused by $F_i$ not reaching sufficiently large values and whenever it increases for parameter values beyond the maximum of $F$, $F_o$ decreases even faster. Increasing $\beta$ does not lead to greatly improved results.
\\
Moving to $\alpha =3$, from where on $\mu$ is a mass scale, we again do not find any point in the parameter space that allows for $F\approx 5$, but can at least reach $F\sim 10^{-117}$.
\\
The first model we find that allows for $F\approx 5$ is the $(5,7)$-symmetron around the parameter space point $(\Lambda,\mathcal{M},\mu) = (10^{37.4},10^{58.7},10^{24.0})\,\text{eV}$. Interestingly, from $\alpha =5$ on, changing $\beta$ can actually have a significant impact on the results. However, while for $\alpha = 6$ we still need at least $\beta = \alpha +2$, from $\alpha = 7$ on we can obtain $F\approx 5$ even for $\beta = \alpha +1$. 
\\
With increasing $\alpha$ we have to use smaller $\Lambda$ in order to reach $F\approx 5$, such that we are getting closer to the Planck scale. For example, in the $(11,12)$-symmetron model we find $F\approx 5$ at the point $(\Lambda,\mathcal{M},\mu) = (10^{27.3},10^{58.7},10^{24.0})\,\text{eV}$, where $\Lambda$ is close to $M_P$.
\\
As can be seen in the examples presented above, increasing $\alpha$ is useful for reducing $\Lambda$, but $\mathcal{M}$ remains more than $30$ order of magnitude above the Planck scale independent of the choice of $(\alpha,\beta)$.


\section{Conclusions}
\label{sec:Conclusion}

In this article we discussed generalizations of the symmetron model, characterized by a pair of positive integers $(\alpha,\beta)$, and investigated for what parameter values their fifth forces can explain the difference between baryonic and lens masses of galaxies, which is otherwise attributed to particle DM. For this, we first reviewed generalized symmetrons and then derived a measure $F$ that allowed us to compare the lensing contribution from Newtonian gravity with one from a gravity-like fifth force induced by a disformally coupling generalized symmetron.
\\
We looked at a Milky Way-like galaxy and checked for a selection of generalized symmetron models for what parameter space values $F \approx 5$, which corresponds to the expected ratio between DM and baryonic matter in galaxies, is fulfilled. With $B=10^{-49}\,\text{eV}^{-4}$ we chose a disformal coupling parameter close to the maximal value allowed by current experimental constraints.  For the standard symmetron, corresponding to $(\alpha,\beta) = (1,2)$, we found that a tiny self-coupling constant $\lambda$ and a matter coupling mass scale $\mathcal{M}$ more than $30$ orders of magnitude larger than the Planck mass $M_P$ are required in order to find $F \approx 5$. Increasing $\beta$, which required us to work with a mass scale $\Lambda$ instead of a dimensionless $\lambda$, led to both $\Lambda$ and $\mathcal{M}$ being at least $30$ orders of magnitude above the Planck scale. Finally, also varying $\alpha$ was most promising since it allowed us to use generalized symmetron fifth forces as explanations for the observed lensing excess at mass scales $\Lambda$ and $\mu$ around or even below the Planck scale. However, in no model it was possible to significantly reduce the value for $\mathcal{M}$, which means it must always be much larger than $M_P$. 
\\
From Refs.\,\,\cite{Burrage2018Sym} and \cite{Kading2019} we know that the $(1,2)$-symmetron is not able to successfully act as an alternative to particle DM since it cannot explain lensing. This is also reflected in the fact that, as we showed in the present article, the standard symmetron fifth force would require an extremely small $\lambda$ for $F \approx 5$. In contrast, some generalized symmetron models with larger $\alpha$ are in so far better in explaining the observed lensing as they instead require the self-interaction mass parameter $\Lambda$ to be around the Planck scale, which is a typical value expected for many screened scalar field models. Though, every model, even for very large $\alpha$ and $\beta$, required $\mathcal{M}$ to be at least $30$ orders of magnitude above $M_P$ in order to reach $F \approx 5$. There are several possible ways around this conundrum: either we accept that we have such a large fundamental mass scale in Nature, we relax the requirement on $F$ and want a generalized symmetron fifth force to only partially explain DM, we find a way to relax the requirement $\varphi \ll \mathcal{M}$, or we consider a hybrid model between modified gravity and particle DM, as was suggested for the standard symmetron in Ref.\,\,\cite{Burrage2018Sym}.
\\
In any case, generalized symmetrons beyond $(\alpha,\beta) =(1,2)$ are interesting theories to study since, to date, there are no experimental constraints on their parameter spaces. In addition, redoing the analyses made in Refs.\,\,\cite{Burrage2016_2} and \cite{OHare:2018ayv} for some of the more promising models, for example theories like $(\alpha,\beta) =(5,7)$ and beyond, might in the future demonstrate that generalized symmetrons are actually good alternatives to particle DM.


\begin{acknowledgments}
CK is grateful to C. Burrage and M. Pitschmann for useful discussions, and to C. Voith and B. Burazor Domazet for spotting a sign mistake in the lensing calculation. This article was supported by the Austrian Science Fund (FWF): P 34240-N, and is based upon work from COST Action COSMIC WISPers CA21106,
supported by COST (European Cooperation in Science and Technology).
\end{acknowledgments}


\bibliography{Lensing}

\begin{thebibliography}{69}%
\makeatletter
\providecommand \@ifxundefined [1]{%
 \@ifx{#1\undefined}
}%
\providecommand \@ifnum [1]{%
 \ifnum #1\expandafter \@firstoftwo
 \else \expandafter \@secondoftwo
 \fi
}%
\providecommand \@ifx [1]{%
 \ifx #1\expandafter \@firstoftwo
 \else \expandafter \@secondoftwo
 \fi
}%
\providecommand \natexlab [1]{#1}%
\providecommand \enquote  [1]{``#1''}%
\providecommand \bibnamefont  [1]{#1}%
\providecommand \bibfnamefont [1]{#1}%
\providecommand \citenamefont [1]{#1}%
\providecommand \href@noop [0]{\@secondoftwo}%
\providecommand \href [0]{\begingroup \@sanitize@url \@href}%
\providecommand \@href[1]{\@@startlink{#1}\@@href}%
\providecommand \@@href[1]{\endgroup#1\@@endlink}%
\providecommand \@sanitize@url [0]{\catcode `\\12\catcode `\$12\catcode
  `\&12\catcode `\#12\catcode `\^12\catcode `\_12\catcode `\%12\relax}%
\providecommand \@@startlink[1]{}%
\providecommand \@@endlink[0]{}%
\providecommand \url  [0]{\begingroup\@sanitize@url \@url }%
\providecommand \@url [1]{\endgroup\@href {#1}{\urlprefix }}%
\providecommand \urlprefix  [0]{URL }%
\providecommand \Eprint [0]{\href }%
\providecommand \doibase [0]{https://doi.org/}%
\providecommand \selectlanguage [0]{\@gobble}%
\providecommand \bibinfo  [0]{\@secondoftwo}%
\providecommand \bibfield  [0]{\@secondoftwo}%
\providecommand \translation [1]{[#1]}%
\providecommand \BibitemOpen [0]{}%
\providecommand \bibitemStop [0]{}%
\providecommand \bibitemNoStop [0]{.\EOS\space}%
\providecommand \EOS [0]{\spacefactor3000\relax}%
\providecommand \BibitemShut  [1]{\csname bibitem#1\endcsname}%
\let\auto@bib@innerbib\@empty
\bibitem [{\citenamefont {{Fujii, Yasunori and Maeda,
  Kei-ichi}}(2003)}]{Fujii2003}%
  \BibitemOpen
  \bibfield  {author} {\bibinfo {author} {\bibnamefont {{Fujii, Yasunori and
  Maeda, Kei-ichi}}},\ }\href {https://doi.org/10.1017/CBO9780511535093} {\emph
  {\bibinfo {title} {The Scalar-Tensor Theory of Gravitation}}},\ Cambridge
  Monographs on Mathematical Physics\ (\bibinfo  {publisher} {Cambridge
  University Press},\ \bibinfo {year} {2003})\BibitemShut {NoStop}%
\bibitem [{\citenamefont {Clifton}\ \emph {et~al.}(2012)\citenamefont
  {Clifton}, \citenamefont {Ferreira}, \citenamefont {Padilla},\ and\
  \citenamefont {Skordis}}]{Clifton2011}%
  \BibitemOpen
  \bibfield  {author} {\bibinfo {author} {\bibfnamefont {T.}~\bibnamefont
  {Clifton}}, \bibinfo {author} {\bibfnamefont {P.~G.}\ \bibnamefont
  {Ferreira}}, \bibinfo {author} {\bibfnamefont {A.}~\bibnamefont {Padilla}},\
  and\ \bibinfo {author} {\bibfnamefont {C.}~\bibnamefont {Skordis}},\
  }\bibfield  {title} {\bibinfo {title} {Modified gravity and cosmology},\
  }\href {https://doi.org/https://doi.org/10.1016/j.physrep.2012.01.001}
  {\bibfield  {journal} {\bibinfo  {journal} {Physics Reports}\ }\textbf
  {\bibinfo {volume} {513}},\ \bibinfo {pages} {1} (\bibinfo {year} {2012})},\
  \bibinfo {note} {modified Gravity and Cosmology}\BibitemShut {NoStop}%
\bibitem [{\citenamefont {Joyce}\ \emph {et~al.}(2015)\citenamefont {Joyce},
  \citenamefont {Jain}, \citenamefont {Khoury},\ and\ \citenamefont
  {Trodden}}]{Joyce2014}%
  \BibitemOpen
  \bibfield  {author} {\bibinfo {author} {\bibfnamefont {A.}~\bibnamefont
  {Joyce}}, \bibinfo {author} {\bibfnamefont {B.}~\bibnamefont {Jain}},
  \bibinfo {author} {\bibfnamefont {J.}~\bibnamefont {Khoury}},\ and\ \bibinfo
  {author} {\bibfnamefont {M.}~\bibnamefont {Trodden}},\ }\bibfield  {title}
  {\bibinfo {title} {{Beyond the Cosmological Standard Model}},\ }\href
  {https://doi.org/10.1016/j.physrep.2014.12.002} {\bibfield  {journal}
  {\bibinfo  {journal} {Phys. Rept.}\ }\textbf {\bibinfo {volume} {568}},\
  \bibinfo {pages} {1} (\bibinfo {year} {2015})},\ \Eprint
  {https://arxiv.org/abs/1407.0059} {arXiv:1407.0059 [astro-ph.CO]}
  \BibitemShut {NoStop}%
\bibitem [{\citenamefont {Dickey}\ \emph {et~al.}(1994)\citenamefont {Dickey},
  \citenamefont {Bender}, \citenamefont {Faller}, \citenamefont {Newhall},
  \citenamefont {Ricklefs}, \citenamefont {Ries}, \citenamefont {Shelus},
  \citenamefont {Veillet}, \citenamefont {Whipple}, \citenamefont {Wiant},
  \citenamefont {Williams},\ and\ \citenamefont {Yoder}}]{Dickey1994}%
  \BibitemOpen
  \bibfield  {author} {\bibinfo {author} {\bibfnamefont {J.~O.}\ \bibnamefont
  {Dickey}}, \bibinfo {author} {\bibfnamefont {P.~L.}\ \bibnamefont {Bender}},
  \bibinfo {author} {\bibfnamefont {J.~E.}\ \bibnamefont {Faller}}, \bibinfo
  {author} {\bibfnamefont {X.~X.}\ \bibnamefont {Newhall}}, \bibinfo {author}
  {\bibfnamefont {R.~L.}\ \bibnamefont {Ricklefs}}, \bibinfo {author}
  {\bibfnamefont {J.~G.}\ \bibnamefont {Ries}}, \bibinfo {author}
  {\bibfnamefont {P.~J.}\ \bibnamefont {Shelus}}, \bibinfo {author}
  {\bibfnamefont {C.}~\bibnamefont {Veillet}}, \bibinfo {author} {\bibfnamefont
  {A.~L.}\ \bibnamefont {Whipple}}, \bibinfo {author} {\bibfnamefont {J.~R.}\
  \bibnamefont {Wiant}}, \bibinfo {author} {\bibfnamefont {J.~G.}\ \bibnamefont
  {Williams}},\ and\ \bibinfo {author} {\bibfnamefont {C.~F.}\ \bibnamefont
  {Yoder}},\ }\bibfield  {title} {\bibinfo {title} {{Lunar Laser Ranging: A
  Continuing Legacy of the Apollo Program}},\ }\href
  {https://doi.org/10.1126/science.265.5171.482} {\bibfield  {journal}
  {\bibinfo  {journal} {Science}\ }\textbf {\bibinfo {volume} {265}},\ \bibinfo
  {pages} {482} (\bibinfo {year} {1994})}\BibitemShut {NoStop}%
\bibitem [{\citenamefont {Adelberger}\ \emph {et~al.}(2003)\citenamefont
  {Adelberger}, \citenamefont {Heckel},\ and\ \citenamefont
  {Nelson}}]{Adelberger2003}%
  \BibitemOpen
  \bibfield  {author} {\bibinfo {author} {\bibfnamefont {E.}~\bibnamefont
  {Adelberger}}, \bibinfo {author} {\bibfnamefont {B.}~\bibnamefont {Heckel}},\
  and\ \bibinfo {author} {\bibfnamefont {A.}~\bibnamefont {Nelson}},\
  }\bibfield  {title} {\bibinfo {title} {{Tests of the Gravitational
  Inverse-Square Law}},\ }\href
  {https://doi.org/10.1146/annurev.nucl.53.041002.110503} {\bibfield  {journal}
  {\bibinfo  {journal} {Annual Review of Nuclear and Particle Science}\
  }\textbf {\bibinfo {volume} {53}},\ \bibinfo {pages} {77} (\bibinfo {year}
  {2003})}\BibitemShut {NoStop}%
\bibitem [{\citenamefont {Kapner}\ \emph {et~al.}(2007)\citenamefont {Kapner},
  \citenamefont {Cook}, \citenamefont {Adelberger}, \citenamefont {Gundlach},
  \citenamefont {Heckel}, \citenamefont {Hoyle},\ and\ \citenamefont
  {Swanson}}]{Kapner2007}%
  \BibitemOpen
  \bibfield  {author} {\bibinfo {author} {\bibfnamefont {D.~J.}\ \bibnamefont
  {Kapner}}, \bibinfo {author} {\bibfnamefont {T.~S.}\ \bibnamefont {Cook}},
  \bibinfo {author} {\bibfnamefont {E.~G.}\ \bibnamefont {Adelberger}},
  \bibinfo {author} {\bibfnamefont {J.~H.}\ \bibnamefont {Gundlach}}, \bibinfo
  {author} {\bibfnamefont {B.~R.}\ \bibnamefont {Heckel}}, \bibinfo {author}
  {\bibfnamefont {C.~D.}\ \bibnamefont {Hoyle}},\ and\ \bibinfo {author}
  {\bibfnamefont {H.~E.}\ \bibnamefont {Swanson}},\ }\bibfield  {title}
  {\bibinfo {title} {{Tests of the Gravitational Inverse-Square Law below the
  Dark-Energy Length Scale}},\ }\href
  {https://doi.org/10.1103/PhysRevLett.98.021101} {\bibfield  {journal}
  {\bibinfo  {journal} {Phys. Rev. Lett.}\ }\textbf {\bibinfo {volume} {98}},\
  \bibinfo {pages} {021101} (\bibinfo {year} {2007})}\BibitemShut {NoStop}%
\bibitem [{\citenamefont {Ishak}(2019)}]{Ishak2018}%
  \BibitemOpen
  \bibfield  {author} {\bibinfo {author} {\bibfnamefont {M.}~\bibnamefont
  {Ishak}},\ }\bibfield  {title} {\bibinfo {title} {{Testing General Relativity
  in Cosmology}},\ }\href {https://doi.org/10.1007/s41114-018-0017-4}
  {\bibfield  {journal} {\bibinfo  {journal} {Living Rev. Rel.}\ }\textbf
  {\bibinfo {volume} {22}},\ \bibinfo {pages} {1} (\bibinfo {year} {2019})},\
  \Eprint {https://arxiv.org/abs/1806.10122} {arXiv:1806.10122 [astro-ph.CO]}
  \BibitemShut {NoStop}%
\bibitem [{\citenamefont {Burrage}\ and\ \citenamefont
  {Sakstein}(2018)}]{BurrageSak}%
  \BibitemOpen
  \bibfield  {author} {\bibinfo {author} {\bibfnamefont {C.}~\bibnamefont
  {Burrage}}\ and\ \bibinfo {author} {\bibfnamefont {J.}~\bibnamefont
  {Sakstein}},\ }\bibfield  {title} {\bibinfo {title} {{Tests of Chameleon
  Gravity}},\ }\href {https://doi.org/10.1007/s41114-018-0011-x} {\bibfield
  {journal} {\bibinfo  {journal} {Living Rev. Rel.}\ }\textbf {\bibinfo
  {volume} {21}},\ \bibinfo {pages} {1} (\bibinfo {year} {2018})},\ \Eprint
  {https://arxiv.org/abs/1709.09071} {arXiv:1709.09071 [astro-ph.CO]}
  \BibitemShut {NoStop}%
\bibitem [{\citenamefont {Brax}\ \emph {et~al.}(2021)\citenamefont {Brax},
  \citenamefont {Casas}, \citenamefont {Desmond},\ and\ \citenamefont
  {Elder}}]{Brax:2021wcv}%
  \BibitemOpen
  \bibfield  {author} {\bibinfo {author} {\bibfnamefont {P.}~\bibnamefont
  {Brax}}, \bibinfo {author} {\bibfnamefont {S.}~\bibnamefont {Casas}},
  \bibinfo {author} {\bibfnamefont {H.}~\bibnamefont {Desmond}},\ and\ \bibinfo
  {author} {\bibfnamefont {B.}~\bibnamefont {Elder}},\ }\bibfield  {title}
  {\bibinfo {title} {{Testing Screened Modified Gravity}},\ }\href
  {https://doi.org/10.3390/universe8010011} {\bibfield  {journal} {\bibinfo
  {journal} {Universe}\ }\textbf {\bibinfo {volume} {8}},\ \bibinfo {pages}
  {11} (\bibinfo {year} {2021})},\ \Eprint {https://arxiv.org/abs/2201.10817}
  {arXiv:2201.10817 [gr-qc]} \BibitemShut {NoStop}%
\bibitem [{\citenamefont {Khoury}\ and\ \citenamefont
  {Weltman}(2004{\natexlab{a}})}]{Khoury2003}%
  \BibitemOpen
  \bibfield  {author} {\bibinfo {author} {\bibfnamefont {J.}~\bibnamefont
  {Khoury}}\ and\ \bibinfo {author} {\bibfnamefont {A.}~\bibnamefont
  {Weltman}},\ }\bibfield  {title} {\bibinfo {title} {{Chameleon cosmology}},\
  }\href {https://doi.org/10.1103/PhysRevD.69.044026} {\bibfield  {journal}
  {\bibinfo  {journal} {Phys. Rev. D}\ }\textbf {\bibinfo {volume} {69}},\
  \bibinfo {pages} {044026} (\bibinfo {year} {2004}{\natexlab{a}})},\ \Eprint
  {https://arxiv.org/abs/astro-ph/0309411} {arXiv:astro-ph/0309411}
  \BibitemShut {NoStop}%
\bibitem [{\citenamefont {Khoury}\ and\ \citenamefont
  {Weltman}(2004{\natexlab{b}})}]{Khoury20032}%
  \BibitemOpen
  \bibfield  {author} {\bibinfo {author} {\bibfnamefont {J.}~\bibnamefont
  {Khoury}}\ and\ \bibinfo {author} {\bibfnamefont {A.}~\bibnamefont
  {Weltman}},\ }\bibfield  {title} {\bibinfo {title} {{Chameleon fields:
  Awaiting surprises for tests of gravity in space}},\ }\href
  {https://doi.org/10.1103/PhysRevLett.93.171104} {\bibfield  {journal}
  {\bibinfo  {journal} {Phys. Rev. Lett.}\ }\textbf {\bibinfo {volume} {93}},\
  \bibinfo {pages} {171104} (\bibinfo {year} {2004}{\natexlab{b}})},\ \Eprint
  {https://arxiv.org/abs/astro-ph/0309300} {arXiv:astro-ph/0309300}
  \BibitemShut {NoStop}%
\bibitem [{\citenamefont {Dehnen}\ \emph {et~al.}(1992)\citenamefont {Dehnen},
  \citenamefont {Frommert},\ and\ \citenamefont {Ghaboussi}}]{Dehnen1992}%
  \BibitemOpen
  \bibfield  {author} {\bibinfo {author} {\bibfnamefont {H.}~\bibnamefont
  {Dehnen}}, \bibinfo {author} {\bibfnamefont {H.}~\bibnamefont {Frommert}},\
  and\ \bibinfo {author} {\bibfnamefont {F.}~\bibnamefont {Ghaboussi}},\
  }\bibfield  {title} {\bibinfo {title} {{Higgs field and a new scalar - tensor
  theory of gravity}},\ }\href {https://doi.org/10.1007/BF00674344} {\bibfield
  {journal} {\bibinfo  {journal} {Int. J. Theor. Phys.}\ }\textbf {\bibinfo
  {volume} {31}},\ \bibinfo {pages} {109} (\bibinfo {year} {1992})}\BibitemShut
  {NoStop}%
\bibitem [{\citenamefont {Gessner}(1992)}]{Gessner1992}%
  \BibitemOpen
  \bibfield  {author} {\bibinfo {author} {\bibfnamefont {E.}~\bibnamefont
  {Gessner}},\ }\bibfield  {title} {\bibinfo {title} {{A new scalar tensor
  theory for gravity and the flat rotation curves of spiral galaxies}},\ }\href
  {https://doi.org/10.1007/BF00645239} {\bibfield  {journal} {\bibinfo
  {journal} {Astrophys. Space Sci.}\ }\textbf {\bibinfo {volume} {196}},\
  \bibinfo {pages} {29} (\bibinfo {year} {1992})}\BibitemShut {NoStop}%
\bibitem [{\citenamefont {Damour}\ and\ \citenamefont
  {Polyakov}(1994)}]{Damour1994}%
  \BibitemOpen
  \bibfield  {author} {\bibinfo {author} {\bibfnamefont {T.}~\bibnamefont
  {Damour}}\ and\ \bibinfo {author} {\bibfnamefont {A.~M.}\ \bibnamefont
  {Polyakov}},\ }\bibfield  {title} {\bibinfo {title} {{The String dilaton and
  a least coupling principle}},\ }\href
  {https://doi.org/10.1016/0550-3213(94)90143-0} {\bibfield  {journal}
  {\bibinfo  {journal} {Nucl. Phys. B}\ }\textbf {\bibinfo {volume} {423}},\
  \bibinfo {pages} {532} (\bibinfo {year} {1994})},\ \Eprint
  {https://arxiv.org/abs/hep-th/9401069} {arXiv:hep-th/9401069} \BibitemShut
  {NoStop}%
\bibitem [{\citenamefont {Pietroni}(2005)}]{Pietroni2005}%
  \BibitemOpen
  \bibfield  {author} {\bibinfo {author} {\bibfnamefont {M.}~\bibnamefont
  {Pietroni}},\ }\bibfield  {title} {\bibinfo {title} {Dark energy
  condensation},\ }\href {https://doi.org/10.1103/PhysRevD.72.043535}
  {\bibfield  {journal} {\bibinfo  {journal} {Phys. Rev. D}\ }\textbf {\bibinfo
  {volume} {72}},\ \bibinfo {pages} {043535} (\bibinfo {year}
  {2005})}\BibitemShut {NoStop}%
\bibitem [{\citenamefont {Olive}\ and\ \citenamefont
  {Pospelov}(2008)}]{Olive2008}%
  \BibitemOpen
  \bibfield  {author} {\bibinfo {author} {\bibfnamefont {K.~A.}\ \bibnamefont
  {Olive}}\ and\ \bibinfo {author} {\bibfnamefont {M.}~\bibnamefont
  {Pospelov}},\ }\bibfield  {title} {\bibinfo {title} {Environmental dependence
  of masses and coupling constants},\ }\href
  {https://doi.org/10.1103/PhysRevD.77.043524} {\bibfield  {journal} {\bibinfo
  {journal} {Phys. Rev. D}\ }\textbf {\bibinfo {volume} {77}},\ \bibinfo
  {pages} {043524} (\bibinfo {year} {2008})}\BibitemShut {NoStop}%
\bibitem [{\citenamefont {Brax}\ \emph {et~al.}(2010)\citenamefont {Brax},
  \citenamefont {van~de Bruck}, \citenamefont {Davis},\ and\ \citenamefont
  {Shaw}}]{Brax2010}%
  \BibitemOpen
  \bibfield  {author} {\bibinfo {author} {\bibfnamefont {P.}~\bibnamefont
  {Brax}}, \bibinfo {author} {\bibfnamefont {C.}~\bibnamefont {van~de Bruck}},
  \bibinfo {author} {\bibfnamefont {A.-C.}\ \bibnamefont {Davis}},\ and\
  \bibinfo {author} {\bibfnamefont {D.}~\bibnamefont {Shaw}},\ }\bibfield
  {title} {\bibinfo {title} {Dilaton and modified gravity},\ }\href
  {https://doi.org/10.1103/PhysRevD.82.063519} {\bibfield  {journal} {\bibinfo
  {journal} {Phys. Rev. D}\ }\textbf {\bibinfo {volume} {82}},\ \bibinfo
  {pages} {063519} (\bibinfo {year} {2010})}\BibitemShut {NoStop}%
\bibitem [{\citenamefont {Hinterbichler}\ and\ \citenamefont
  {Khoury}(2010)}]{Hinterbichler2010}%
  \BibitemOpen
  \bibfield  {author} {\bibinfo {author} {\bibfnamefont {K.}~\bibnamefont
  {Hinterbichler}}\ and\ \bibinfo {author} {\bibfnamefont {J.}~\bibnamefont
  {Khoury}},\ }\bibfield  {title} {\bibinfo {title} {{Symmetron Fields:
  Screening Long-Range Forces Through Local Symmetry Restoration}},\ }\href
  {https://doi.org/10.1103/PhysRevLett.104.231301} {\bibfield  {journal}
  {\bibinfo  {journal} {Phys. Rev. Lett.}\ }\textbf {\bibinfo {volume} {104}},\
  \bibinfo {pages} {231301} (\bibinfo {year} {2010})},\ \Eprint
  {https://arxiv.org/abs/1001.4525} {arXiv:1001.4525 [hep-th]} \BibitemShut
  {NoStop}%
\bibitem [{\citenamefont {Hinterbichler}\ \emph {et~al.}(2011)\citenamefont
  {Hinterbichler}, \citenamefont {Khoury}, \citenamefont {Levy},\ and\
  \citenamefont {Matas}}]{Hinterbichler2011}%
  \BibitemOpen
  \bibfield  {author} {\bibinfo {author} {\bibfnamefont {K.}~\bibnamefont
  {Hinterbichler}}, \bibinfo {author} {\bibfnamefont {J.}~\bibnamefont
  {Khoury}}, \bibinfo {author} {\bibfnamefont {A.}~\bibnamefont {Levy}},\ and\
  \bibinfo {author} {\bibfnamefont {A.}~\bibnamefont {Matas}},\ }\bibfield
  {title} {\bibinfo {title} {{Symmetron Cosmology}},\ }\href
  {https://doi.org/10.1103/PhysRevD.84.103521} {\bibfield  {journal} {\bibinfo
  {journal} {Phys. Rev. D}\ }\textbf {\bibinfo {volume} {84}},\ \bibinfo
  {pages} {103521} (\bibinfo {year} {2011})},\ \Eprint
  {https://arxiv.org/abs/1107.2112} {arXiv:1107.2112 [astro-ph.CO]}
  \BibitemShut {NoStop}%
\bibitem [{\citenamefont {Gasperini}\ \emph {et~al.}(2002)\citenamefont
  {Gasperini}, \citenamefont {Piazza},\ and\ \citenamefont
  {Veneziano}}]{Gasperini:2001pc}%
  \BibitemOpen
  \bibfield  {author} {\bibinfo {author} {\bibfnamefont {M.}~\bibnamefont
  {Gasperini}}, \bibinfo {author} {\bibfnamefont {F.}~\bibnamefont {Piazza}},\
  and\ \bibinfo {author} {\bibfnamefont {G.}~\bibnamefont {Veneziano}},\
  }\bibfield  {title} {\bibinfo {title} {{Quintessence as a runaway dilaton}},\
  }\href {https://doi.org/10.1103/PhysRevD.65.023508} {\bibfield  {journal}
  {\bibinfo  {journal} {Phys. Rev. D}\ }\textbf {\bibinfo {volume} {65}},\
  \bibinfo {pages} {023508} (\bibinfo {year} {2002})},\ \Eprint
  {https://arxiv.org/abs/gr-qc/0108016} {arXiv:gr-qc/0108016} \BibitemShut
  {NoStop}%
\bibitem [{\citenamefont {Damour}\ \emph
  {et~al.}(2002{\natexlab{a}})\citenamefont {Damour}, \citenamefont {Piazza},\
  and\ \citenamefont {Veneziano}}]{Damour:2002nv}%
  \BibitemOpen
  \bibfield  {author} {\bibinfo {author} {\bibfnamefont {T.}~\bibnamefont
  {Damour}}, \bibinfo {author} {\bibfnamefont {F.}~\bibnamefont {Piazza}},\
  and\ \bibinfo {author} {\bibfnamefont {G.}~\bibnamefont {Veneziano}},\
  }\bibfield  {title} {\bibinfo {title} {{Violations of the equivalence
  principle in a dilaton runaway scenario}},\ }\href
  {https://doi.org/10.1103/PhysRevD.66.046007} {\bibfield  {journal} {\bibinfo
  {journal} {Phys. Rev. D}\ }\textbf {\bibinfo {volume} {66}},\ \bibinfo
  {pages} {046007} (\bibinfo {year} {2002}{\natexlab{a}})},\ \Eprint
  {https://arxiv.org/abs/hep-th/0205111} {arXiv:hep-th/0205111} \BibitemShut
  {NoStop}%
\bibitem [{\citenamefont {Damour}\ \emph
  {et~al.}(2002{\natexlab{b}})\citenamefont {Damour}, \citenamefont {Piazza},\
  and\ \citenamefont {Veneziano}}]{Damour:2002mi}%
  \BibitemOpen
  \bibfield  {author} {\bibinfo {author} {\bibfnamefont {T.}~\bibnamefont
  {Damour}}, \bibinfo {author} {\bibfnamefont {F.}~\bibnamefont {Piazza}},\
  and\ \bibinfo {author} {\bibfnamefont {G.}~\bibnamefont {Veneziano}},\
  }\bibfield  {title} {\bibinfo {title} {{Runaway dilaton and equivalence
  principle violations}},\ }\href
  {https://doi.org/10.1103/PhysRevLett.89.081601} {\bibfield  {journal}
  {\bibinfo  {journal} {Phys. Rev. Lett.}\ }\textbf {\bibinfo {volume} {89}},\
  \bibinfo {pages} {081601} (\bibinfo {year} {2002}{\natexlab{b}})},\ \Eprint
  {https://arxiv.org/abs/gr-qc/0204094} {arXiv:gr-qc/0204094} \BibitemShut
  {NoStop}%
\bibitem [{\citenamefont {Brax}\ \emph {et~al.}(2011)\citenamefont {Brax},
  \citenamefont {van~de Bruck}, \citenamefont {Davis}, \citenamefont {Li},\
  and\ \citenamefont {Shaw}}]{Brax:2011ja}%
  \BibitemOpen
  \bibfield  {author} {\bibinfo {author} {\bibfnamefont {P.}~\bibnamefont
  {Brax}}, \bibinfo {author} {\bibfnamefont {C.}~\bibnamefont {van~de Bruck}},
  \bibinfo {author} {\bibfnamefont {A.-C.}\ \bibnamefont {Davis}}, \bibinfo
  {author} {\bibfnamefont {B.}~\bibnamefont {Li}},\ and\ \bibinfo {author}
  {\bibfnamefont {D.~J.}\ \bibnamefont {Shaw}},\ }\bibfield  {title} {\bibinfo
  {title} {{Nonlinear Structure Formation with the Environmentally Dependent
  Dilaton}},\ }\href {https://doi.org/10.1103/PhysRevD.83.104026} {\bibfield
  {journal} {\bibinfo  {journal} {Phys. Rev. D}\ }\textbf {\bibinfo {volume}
  {83}},\ \bibinfo {pages} {104026} (\bibinfo {year} {2011})},\ \Eprint
  {https://arxiv.org/abs/1102.3692} {arXiv:1102.3692 [astro-ph.CO]}
  \BibitemShut {NoStop}%
\bibitem [{\citenamefont {Brax}\ \emph {et~al.}(2022)\citenamefont {Brax},
  \citenamefont {Fischer}, \citenamefont {K\"ading},\ and\ \citenamefont
  {Pitschmann}}]{Brax2022}%
  \BibitemOpen
  \bibfield  {author} {\bibinfo {author} {\bibfnamefont {P.}~\bibnamefont
  {Brax}}, \bibinfo {author} {\bibfnamefont {H.}~\bibnamefont {Fischer}},
  \bibinfo {author} {\bibfnamefont {C.}~\bibnamefont {K\"ading}},\ and\
  \bibinfo {author} {\bibfnamefont {M.}~\bibnamefont {Pitschmann}},\ }\bibfield
   {title} {\bibinfo {title} {{The environment dependent dilaton in the
  laboratory and the solar system}},\ }\href
  {https://doi.org/10.1140/epjc/s10052-022-10905-w} {\bibfield  {journal}
  {\bibinfo  {journal} {Eur. Phys. J. C}\ }\textbf {\bibinfo {volume} {82}},\
  \bibinfo {pages} {934} (\bibinfo {year} {2022})},\ \Eprint
  {https://arxiv.org/abs/2203.12512} {arXiv:2203.12512 [gr-qc]} \BibitemShut
  {NoStop}%
\bibitem [{\citenamefont {Dvali}\ \emph {et~al.}(2000)\citenamefont {Dvali},
  \citenamefont {Gabadadze},\ and\ \citenamefont {Porrati}}]{Dvali2000}%
  \BibitemOpen
  \bibfield  {author} {\bibinfo {author} {\bibfnamefont {G.~R.}\ \bibnamefont
  {Dvali}}, \bibinfo {author} {\bibfnamefont {G.}~\bibnamefont {Gabadadze}},\
  and\ \bibinfo {author} {\bibfnamefont {M.}~\bibnamefont {Porrati}},\
  }\bibfield  {title} {\bibinfo {title} {{4-D gravity on a brane in 5-D
  Minkowski space}},\ }\href {https://doi.org/10.1016/S0370-2693(00)00669-9}
  {\bibfield  {journal} {\bibinfo  {journal} {Phys. Lett. B}\ }\textbf
  {\bibinfo {volume} {485}},\ \bibinfo {pages} {208} (\bibinfo {year}
  {2000})},\ \Eprint {https://arxiv.org/abs/hep-th/0005016}
  {arXiv:hep-th/0005016} \BibitemShut {NoStop}%
\bibitem [{\citenamefont {Nicolis}\ \emph {et~al.}(2009)\citenamefont
  {Nicolis}, \citenamefont {Rattazzi},\ and\ \citenamefont
  {Trincherini}}]{Nicolis2008}%
  \BibitemOpen
  \bibfield  {author} {\bibinfo {author} {\bibfnamefont {A.}~\bibnamefont
  {Nicolis}}, \bibinfo {author} {\bibfnamefont {R.}~\bibnamefont {Rattazzi}},\
  and\ \bibinfo {author} {\bibfnamefont {E.}~\bibnamefont {Trincherini}},\
  }\bibfield  {title} {\bibinfo {title} {{The Galileon as a local modification
  of gravity}},\ }\href {https://doi.org/10.1103/PhysRevD.79.064036} {\bibfield
   {journal} {\bibinfo  {journal} {Phys. Rev. D}\ }\textbf {\bibinfo {volume}
  {79}},\ \bibinfo {pages} {064036} (\bibinfo {year} {2009})},\ \Eprint
  {https://arxiv.org/abs/0811.2197} {arXiv:0811.2197 [hep-th]} \BibitemShut
  {NoStop}%
\bibitem [{\citenamefont {Ali}\ \emph {et~al.}(2012)\citenamefont {Ali},
  \citenamefont {Gannouji}, \citenamefont {Hossain},\ and\ \citenamefont
  {Sami}}]{Ali2012}%
  \BibitemOpen
  \bibfield  {author} {\bibinfo {author} {\bibfnamefont {A.}~\bibnamefont
  {Ali}}, \bibinfo {author} {\bibfnamefont {R.}~\bibnamefont {Gannouji}},
  \bibinfo {author} {\bibfnamefont {M.~W.}\ \bibnamefont {Hossain}},\ and\
  \bibinfo {author} {\bibfnamefont {M.}~\bibnamefont {Sami}},\ }\bibfield
  {title} {\bibinfo {title} {{Light mass galileons: Cosmological dynamics, mass
  screening and observational constraints}},\ }\href
  {https://doi.org/10.1016/j.physletb.2012.10.009} {\bibfield  {journal}
  {\bibinfo  {journal} {Phys. Lett. B}\ }\textbf {\bibinfo {volume} {718}},\
  \bibinfo {pages} {5} (\bibinfo {year} {2012})},\ \Eprint
  {https://arxiv.org/abs/1207.3959} {arXiv:1207.3959 [gr-qc]} \BibitemShut
  {NoStop}%
\bibitem [{\citenamefont {Burrage}\ and\ \citenamefont
  {Sakstein}(2016)}]{Burrage:2016bwy}%
  \BibitemOpen
  \bibfield  {author} {\bibinfo {author} {\bibfnamefont {C.}~\bibnamefont
  {Burrage}}\ and\ \bibinfo {author} {\bibfnamefont {J.}~\bibnamefont
  {Sakstein}},\ }\bibfield  {title} {\bibinfo {title} {{A Compendium of
  Chameleon Constraints}},\ }\href
  {https://doi.org/10.1088/1475-7516/2016/11/045} {\bibfield  {journal}
  {\bibinfo  {journal} {JCAP}\ }\textbf {\bibinfo {volume} {11}},\ \bibinfo
  {pages} {045}},\ \Eprint {https://arxiv.org/abs/1609.01192} {arXiv:1609.01192
  [astro-ph.CO]} \BibitemShut {NoStop}%
\bibitem [{\citenamefont
  {Pokotilovski}(2013{\natexlab{a}})}]{Pokotilovski:2012xuk}%
  \BibitemOpen
  \bibfield  {author} {\bibinfo {author} {\bibfnamefont {Y.~N.}\ \bibnamefont
  {Pokotilovski}},\ }\bibfield  {title} {\bibinfo {title} {{Strongly coupled
  chameleon fields: Possible test with a neutron Lloyd's mirror
  interferometer}},\ }\href {https://doi.org/10.1016/j.physletb.2013.01.022}
  {\bibfield  {journal} {\bibinfo  {journal} {Phys. Lett. B}\ }\textbf
  {\bibinfo {volume} {719}},\ \bibinfo {pages} {341} (\bibinfo {year}
  {2013}{\natexlab{a}})},\ \Eprint {https://arxiv.org/abs/1203.5017}
  {arXiv:1203.5017 [nucl-ex]} \BibitemShut {NoStop}%
\bibitem [{\citenamefont
  {Pokotilovski}(2013{\natexlab{b}})}]{Pokotilovski:2013tma}%
  \BibitemOpen
  \bibfield  {author} {\bibinfo {author} {\bibfnamefont {Y.~N.}\ \bibnamefont
  {Pokotilovski}},\ }\bibfield  {title} {\bibinfo {title} {{Potential of the
  neutron Lloyd`s mirror interferometer for the search for new interactions}},\
  }\href {https://doi.org/10.1134/S106377611309001X} {\bibfield  {journal}
  {\bibinfo  {journal} {J. Exp. Theor. Phys.}\ }\textbf {\bibinfo {volume}
  {116}},\ \bibinfo {pages} {609} (\bibinfo {year} {2013}{\natexlab{b}})},\
  \bibinfo {note} {[Erratum: J.Exp.Theor.Phys. 116, 886 (2013)]},\ \Eprint
  {https://arxiv.org/abs/1311.4679} {arXiv:1311.4679 [nucl-ex]} \BibitemShut
  {NoStop}%
\bibitem [{\citenamefont {Burrage}\ \emph {et~al.}(2015)\citenamefont
  {Burrage}, \citenamefont {Copeland},\ and\ \citenamefont
  {Hinds}}]{Burrage:2014oza}%
  \BibitemOpen
  \bibfield  {author} {\bibinfo {author} {\bibfnamefont {C.}~\bibnamefont
  {Burrage}}, \bibinfo {author} {\bibfnamefont {E.~J.}\ \bibnamefont
  {Copeland}},\ and\ \bibinfo {author} {\bibfnamefont {E.~A.}\ \bibnamefont
  {Hinds}},\ }\bibfield  {title} {\bibinfo {title} {{Probing Dark Energy with
  Atom Interferometry}},\ }\href
  {https://doi.org/10.1088/1475-7516/2015/03/042} {\bibfield  {journal}
  {\bibinfo  {journal} {JCAP}\ }\textbf {\bibinfo {volume} {03}},\ \bibinfo
  {pages} {042}},\ \Eprint {https://arxiv.org/abs/1408.1409} {arXiv:1408.1409
  [astro-ph.CO]} \BibitemShut {NoStop}%
\bibitem [{\citenamefont {Hamilton}\ \emph {et~al.}(2015)\citenamefont
  {Hamilton}, \citenamefont {Jaffe}, \citenamefont {Haslinger}, \citenamefont
  {Simmons}, \citenamefont {M\"uller},\ and\ \citenamefont
  {Khoury}}]{Hamilton:2015zga}%
  \BibitemOpen
  \bibfield  {author} {\bibinfo {author} {\bibfnamefont {P.}~\bibnamefont
  {Hamilton}}, \bibinfo {author} {\bibfnamefont {M.}~\bibnamefont {Jaffe}},
  \bibinfo {author} {\bibfnamefont {P.}~\bibnamefont {Haslinger}}, \bibinfo
  {author} {\bibfnamefont {Q.}~\bibnamefont {Simmons}}, \bibinfo {author}
  {\bibfnamefont {H.}~\bibnamefont {M\"uller}},\ and\ \bibinfo {author}
  {\bibfnamefont {J.}~\bibnamefont {Khoury}},\ }\bibfield  {title} {\bibinfo
  {title} {{Atom-interferometry constraints on dark energy}},\ }\href
  {https://doi.org/10.1126/science.aaa8883} {\bibfield  {journal} {\bibinfo
  {journal} {Science}\ }\textbf {\bibinfo {volume} {349}},\ \bibinfo {pages}
  {849} (\bibinfo {year} {2015})},\ \Eprint {https://arxiv.org/abs/1502.03888}
  {arXiv:1502.03888 [physics.atom-ph]} \BibitemShut {NoStop}%
\bibitem [{\citenamefont {Lemmel}\ \emph {et~al.}(2015)\citenamefont {Lemmel},
  \citenamefont {Brax}, \citenamefont {Ivanov}, \citenamefont {Jenke},
  \citenamefont {Pignol}, \citenamefont {Pitschmann}, \citenamefont {Potocar},
  \citenamefont {Wellenzohn}, \citenamefont {Zawisky},\ and\ \citenamefont
  {Abele}}]{Lemmel:2015kwa}%
  \BibitemOpen
  \bibfield  {author} {\bibinfo {author} {\bibfnamefont {H.}~\bibnamefont
  {Lemmel}}, \bibinfo {author} {\bibfnamefont {P.}~\bibnamefont {Brax}},
  \bibinfo {author} {\bibfnamefont {A.~N.}\ \bibnamefont {Ivanov}}, \bibinfo
  {author} {\bibfnamefont {T.}~\bibnamefont {Jenke}}, \bibinfo {author}
  {\bibfnamefont {G.}~\bibnamefont {Pignol}}, \bibinfo {author} {\bibfnamefont
  {M.}~\bibnamefont {Pitschmann}}, \bibinfo {author} {\bibfnamefont
  {T.}~\bibnamefont {Potocar}}, \bibinfo {author} {\bibfnamefont
  {M.}~\bibnamefont {Wellenzohn}}, \bibinfo {author} {\bibfnamefont
  {M.}~\bibnamefont {Zawisky}},\ and\ \bibinfo {author} {\bibfnamefont
  {H.}~\bibnamefont {Abele}},\ }\bibfield  {title} {\bibinfo {title} {{Neutron
  Interferometry constrains dark energy chameleon fields}},\ }\href
  {https://doi.org/10.1016/j.physletb.2015.02.063} {\bibfield  {journal}
  {\bibinfo  {journal} {Phys. Lett. B}\ }\textbf {\bibinfo {volume} {743}},\
  \bibinfo {pages} {310} (\bibinfo {year} {2015})},\ \Eprint
  {https://arxiv.org/abs/1502.06023} {arXiv:1502.06023 [hep-ph]} \BibitemShut
  {NoStop}%
\bibitem [{\citenamefont {Burrage}\ and\ \citenamefont
  {Copeland}(2016)}]{Burrage:2015lya}%
  \BibitemOpen
  \bibfield  {author} {\bibinfo {author} {\bibfnamefont {C.}~\bibnamefont
  {Burrage}}\ and\ \bibinfo {author} {\bibfnamefont {E.~J.}\ \bibnamefont
  {Copeland}},\ }\bibfield  {title} {\bibinfo {title} {{Using Atom
  Interferometry to Detect Dark Energy}},\ }\href
  {https://doi.org/10.1080/00107514.2015.1060058} {\bibfield  {journal}
  {\bibinfo  {journal} {Contemp. Phys.}\ }\textbf {\bibinfo {volume} {57}},\
  \bibinfo {pages} {164} (\bibinfo {year} {2016})},\ \Eprint
  {https://arxiv.org/abs/1507.07493} {arXiv:1507.07493 [astro-ph.CO]}
  \BibitemShut {NoStop}%
\bibitem [{\citenamefont {Elder}\ \emph {et~al.}(2016)\citenamefont {Elder},
  \citenamefont {Khoury}, \citenamefont {Haslinger}, \citenamefont {Jaffe},
  \citenamefont {M\"uller},\ and\ \citenamefont {Hamilton}}]{Elder:2016yxm}%
  \BibitemOpen
  \bibfield  {author} {\bibinfo {author} {\bibfnamefont {B.}~\bibnamefont
  {Elder}}, \bibinfo {author} {\bibfnamefont {J.}~\bibnamefont {Khoury}},
  \bibinfo {author} {\bibfnamefont {P.}~\bibnamefont {Haslinger}}, \bibinfo
  {author} {\bibfnamefont {M.}~\bibnamefont {Jaffe}}, \bibinfo {author}
  {\bibfnamefont {H.}~\bibnamefont {M\"uller}},\ and\ \bibinfo {author}
  {\bibfnamefont {P.}~\bibnamefont {Hamilton}},\ }\bibfield  {title} {\bibinfo
  {title} {{Chameleon Dark Energy and Atom Interferometry}},\ }\href
  {https://doi.org/10.1103/PhysRevD.94.044051} {\bibfield  {journal} {\bibinfo
  {journal} {Phys. Rev. D}\ }\textbf {\bibinfo {volume} {94}},\ \bibinfo
  {pages} {044051} (\bibinfo {year} {2016})},\ \Eprint
  {https://arxiv.org/abs/1603.06587} {arXiv:1603.06587 [astro-ph.CO]}
  \BibitemShut {NoStop}%
\bibitem [{\citenamefont {Ivanov}\ \emph {et~al.}(2016)\citenamefont {Ivanov},
  \citenamefont {Cronenberg}, \citenamefont {H\"ollwieser}, \citenamefont
  {Pitschmann}, \citenamefont {Jenke}, \citenamefont {Wellenzohn},\ and\
  \citenamefont {Abele}}]{Ivanov:2016rfs}%
  \BibitemOpen
  \bibfield  {author} {\bibinfo {author} {\bibfnamefont {A.~N.}\ \bibnamefont
  {Ivanov}}, \bibinfo {author} {\bibfnamefont {G.}~\bibnamefont {Cronenberg}},
  \bibinfo {author} {\bibfnamefont {R.}~\bibnamefont {H\"ollwieser}}, \bibinfo
  {author} {\bibfnamefont {M.}~\bibnamefont {Pitschmann}}, \bibinfo {author}
  {\bibfnamefont {T.}~\bibnamefont {Jenke}}, \bibinfo {author} {\bibfnamefont
  {M.}~\bibnamefont {Wellenzohn}},\ and\ \bibinfo {author} {\bibfnamefont
  {H.}~\bibnamefont {Abele}},\ }\bibfield  {title} {\bibinfo {title} {{Exact
  solution for chameleon field, self-coupled through the Ratra-Peebles
  potential with $n=1$ and confined between two parallel plates}},\ }\href
  {https://doi.org/10.1103/PhysRevD.94.085005} {\bibfield  {journal} {\bibinfo
  {journal} {Phys. Rev. D}\ }\textbf {\bibinfo {volume} {94}},\ \bibinfo
  {pages} {085005} (\bibinfo {year} {2016})},\ \Eprint
  {https://arxiv.org/abs/1606.06867} {arXiv:1606.06867 [gr-qc]} \BibitemShut
  {NoStop}%
\bibitem [{\citenamefont {Burrage}\ \emph
  {et~al.}(2016{\natexlab{a}})\citenamefont {Burrage}, \citenamefont
  {Kuribayashi-Coleman}, \citenamefont {Stevenson},\ and\ \citenamefont
  {Thrussell}}]{Burrage:2016rkv}%
  \BibitemOpen
  \bibfield  {author} {\bibinfo {author} {\bibfnamefont {C.}~\bibnamefont
  {Burrage}}, \bibinfo {author} {\bibfnamefont {A.}~\bibnamefont
  {Kuribayashi-Coleman}}, \bibinfo {author} {\bibfnamefont {J.}~\bibnamefont
  {Stevenson}},\ and\ \bibinfo {author} {\bibfnamefont {B.}~\bibnamefont
  {Thrussell}},\ }\bibfield  {title} {\bibinfo {title} {{Constraining symmetron
  fields with atom interferometry}},\ }\href
  {https://doi.org/10.1088/1475-7516/2016/12/041} {\bibfield  {journal}
  {\bibinfo  {journal} {JCAP}\ }\textbf {\bibinfo {volume} {12}},\ \bibinfo
  {pages} {041}},\ \Eprint {https://arxiv.org/abs/1609.09275} {arXiv:1609.09275
  [astro-ph.CO]} \BibitemShut {NoStop}%
\bibitem [{\citenamefont {Jaffe}\ \emph {et~al.}(2017)\citenamefont {Jaffe},
  \citenamefont {Haslinger}, \citenamefont {Xu}, \citenamefont {Hamilton},
  \citenamefont {Upadhye}, \citenamefont {Elder}, \citenamefont {Khoury},\ and\
  \citenamefont {M\"uller}}]{Jaffe:2016fsh}%
  \BibitemOpen
  \bibfield  {author} {\bibinfo {author} {\bibfnamefont {M.}~\bibnamefont
  {Jaffe}}, \bibinfo {author} {\bibfnamefont {P.}~\bibnamefont {Haslinger}},
  \bibinfo {author} {\bibfnamefont {V.}~\bibnamefont {Xu}}, \bibinfo {author}
  {\bibfnamefont {P.}~\bibnamefont {Hamilton}}, \bibinfo {author}
  {\bibfnamefont {A.}~\bibnamefont {Upadhye}}, \bibinfo {author} {\bibfnamefont
  {B.}~\bibnamefont {Elder}}, \bibinfo {author} {\bibfnamefont
  {J.}~\bibnamefont {Khoury}},\ and\ \bibinfo {author} {\bibfnamefont
  {H.}~\bibnamefont {M\"uller}},\ }\bibfield  {title} {\bibinfo {title}
  {{Testing sub-gravitational forces on atoms from a miniature, in-vacuum
  source mass}},\ }\href {https://doi.org/10.1038/nphys4189} {\bibfield
  {journal} {\bibinfo  {journal} {Nature Phys.}\ }\textbf {\bibinfo {volume}
  {13}},\ \bibinfo {pages} {938} (\bibinfo {year} {2017})},\ \Eprint
  {https://arxiv.org/abs/1612.05171} {arXiv:1612.05171 [physics.atom-ph]}
  \BibitemShut {NoStop}%
\bibitem [{\citenamefont {Brax}\ and\ \citenamefont
  {Pitschmann}(2018)}]{Brax:2017hna}%
  \BibitemOpen
  \bibfield  {author} {\bibinfo {author} {\bibfnamefont {P.}~\bibnamefont
  {Brax}}\ and\ \bibinfo {author} {\bibfnamefont {M.}~\bibnamefont
  {Pitschmann}},\ }\bibfield  {title} {\bibinfo {title} {{Exact solutions to
  nonlinear symmetron theory: One- and two-mirror systems}},\ }\href
  {https://doi.org/10.1103/PhysRevD.97.064015} {\bibfield  {journal} {\bibinfo
  {journal} {Phys. Rev. D}\ }\textbf {\bibinfo {volume} {97}},\ \bibinfo
  {pages} {064015} (\bibinfo {year} {2018})},\ \Eprint
  {https://arxiv.org/abs/1712.09852} {arXiv:1712.09852 [gr-qc]} \BibitemShut
  {NoStop}%
\bibitem [{\citenamefont {Sabulsky}\ \emph {et~al.}(2019)\citenamefont
  {Sabulsky}, \citenamefont {Dutta}, \citenamefont {Hinds}, \citenamefont
  {Elder}, \citenamefont {Burrage},\ and\ \citenamefont
  {Copeland}}]{Sabulsky:2018jma}%
  \BibitemOpen
  \bibfield  {author} {\bibinfo {author} {\bibfnamefont {D.~O.}\ \bibnamefont
  {Sabulsky}}, \bibinfo {author} {\bibfnamefont {I.}~\bibnamefont {Dutta}},
  \bibinfo {author} {\bibfnamefont {E.~A.}\ \bibnamefont {Hinds}}, \bibinfo
  {author} {\bibfnamefont {B.}~\bibnamefont {Elder}}, \bibinfo {author}
  {\bibfnamefont {C.}~\bibnamefont {Burrage}},\ and\ \bibinfo {author}
  {\bibfnamefont {E.~J.}\ \bibnamefont {Copeland}},\ }\bibfield  {title}
  {\bibinfo {title} {{Experiment to detect dark energy forces using atom
  interferometry}},\ }\href {https://doi.org/10.1103/PhysRevLett.123.061102}
  {\bibfield  {journal} {\bibinfo  {journal} {Phys. Rev. Lett.}\ }\textbf
  {\bibinfo {volume} {123}},\ \bibinfo {pages} {061102} (\bibinfo {year}
  {2019})},\ \Eprint {https://arxiv.org/abs/1812.08244} {arXiv:1812.08244
  [physics.atom-ph]} \BibitemShut {NoStop}%
\bibitem [{\citenamefont {Brax}\ \emph {et~al.}(2018)\citenamefont {Brax},
  \citenamefont {Burrage},\ and\ \citenamefont {Davis}}]{Brax:2018iyo}%
  \BibitemOpen
  \bibfield  {author} {\bibinfo {author} {\bibfnamefont {P.}~\bibnamefont
  {Brax}}, \bibinfo {author} {\bibfnamefont {C.}~\bibnamefont {Burrage}},\ and\
  \bibinfo {author} {\bibfnamefont {A.-C.}\ \bibnamefont {Davis}},\ }\bibfield
  {title} {\bibinfo {title} {{Laboratory constraints}},\ }\href
  {https://doi.org/10.1142/S0218271818480097} {\bibfield  {journal} {\bibinfo
  {journal} {Int. J. Mod. Phys. D}\ }\textbf {\bibinfo {volume} {27}},\
  \bibinfo {pages} {1848009} (\bibinfo {year} {2018})}\BibitemShut {NoStop}%
\bibitem [{\citenamefont {Cronenberg}\ \emph {et~al.}(2018)\citenamefont
  {Cronenberg}, \citenamefont {Brax}, \citenamefont {Filter}, \citenamefont
  {Geltenbort}, \citenamefont {Jenke}, \citenamefont {Pignol}, \citenamefont
  {Pitschmann}, \citenamefont {Thalhammer},\ and\ \citenamefont
  {Abele}}]{Cronenberg:2018qxf}%
  \BibitemOpen
  \bibfield  {author} {\bibinfo {author} {\bibfnamefont {G.}~\bibnamefont
  {Cronenberg}}, \bibinfo {author} {\bibfnamefont {P.}~\bibnamefont {Brax}},
  \bibinfo {author} {\bibfnamefont {H.}~\bibnamefont {Filter}}, \bibinfo
  {author} {\bibfnamefont {P.}~\bibnamefont {Geltenbort}}, \bibinfo {author}
  {\bibfnamefont {T.}~\bibnamefont {Jenke}}, \bibinfo {author} {\bibfnamefont
  {G.}~\bibnamefont {Pignol}}, \bibinfo {author} {\bibfnamefont
  {M.}~\bibnamefont {Pitschmann}}, \bibinfo {author} {\bibfnamefont
  {M.}~\bibnamefont {Thalhammer}},\ and\ \bibinfo {author} {\bibfnamefont
  {H.}~\bibnamefont {Abele}},\ }\bibfield  {title} {\bibinfo {title} {{Acoustic
  Rabi oscillations between gravitational quantum states and impact on
  symmetron dark energy}},\ }\href {https://doi.org/10.1038/s41567-018-0205-x}
  {\bibfield  {journal} {\bibinfo  {journal} {Nature Phys.}\ }\textbf {\bibinfo
  {volume} {14}},\ \bibinfo {pages} {1022} (\bibinfo {year} {2018})},\ \Eprint
  {https://arxiv.org/abs/1902.08775} {arXiv:1902.08775 [hep-ph]} \BibitemShut
  {NoStop}%
\bibitem [{\citenamefont {Hartley}\ \emph
  {et~al.}(2019{\natexlab{a}})\citenamefont {Hartley}, \citenamefont
  {K\"ading}, \citenamefont {Howl},\ and\ \citenamefont
  {Fuentes}}]{Hartley2019}%
  \BibitemOpen
  \bibfield  {author} {\bibinfo {author} {\bibfnamefont {D.}~\bibnamefont
  {Hartley}}, \bibinfo {author} {\bibfnamefont {C.}~\bibnamefont {K\"ading}},
  \bibinfo {author} {\bibfnamefont {R.}~\bibnamefont {Howl}},\ and\ \bibinfo
  {author} {\bibfnamefont {I.}~\bibnamefont {Fuentes}},\ }\bibfield  {title}
  {\bibinfo {title} {{Quantum-enhanced screened dark energy detection}},\
  }\href@noop {} {\  (\bibinfo {year} {2019}{\natexlab{a}})},\ \Eprint
  {https://arxiv.org/abs/1909.02272} {arXiv:1909.02272 [gr-qc]} \BibitemShut
  {NoStop}%
\bibitem [{\citenamefont {Pitschmann}(2021)}]{Pitschmann:2020ejb}%
  \BibitemOpen
  \bibfield  {author} {\bibinfo {author} {\bibfnamefont {M.}~\bibnamefont
  {Pitschmann}},\ }\bibfield  {title} {\bibinfo {title} {{Exact solutions to
  nonlinear symmetron theory: One- and two-mirror systems. II.}},\ }\href
  {https://doi.org/10.1103/PhysRevD.103.084013} {\bibfield  {journal} {\bibinfo
   {journal} {Phys. Rev. D}\ }\textbf {\bibinfo {volume} {103}},\ \bibinfo
  {pages} {084013} (\bibinfo {year} {2021})},\ \bibinfo {note} {[Erratum:
  Phys.Rev.D 106, 109902 (2022)]},\ \Eprint {https://arxiv.org/abs/2012.12752}
  {arXiv:2012.12752 [gr-qc]} \BibitemShut {NoStop}%
\bibitem [{\citenamefont {Brax}\ and\ \citenamefont
  {Fichet}(2019)}]{Brax2018quantch}%
  \BibitemOpen
  \bibfield  {author} {\bibinfo {author} {\bibfnamefont {P.}~\bibnamefont
  {Brax}}\ and\ \bibinfo {author} {\bibfnamefont {S.}~\bibnamefont {Fichet}},\
  }\bibfield  {title} {\bibinfo {title} {{Quantum Chameleons}},\ }\href
  {https://doi.org/10.1103/PhysRevD.99.104049} {\bibfield  {journal} {\bibinfo
  {journal} {Phys. Rev. D}\ }\textbf {\bibinfo {volume} {99}},\ \bibinfo
  {pages} {104049} (\bibinfo {year} {2019})},\ \Eprint
  {https://arxiv.org/abs/1809.10166} {arXiv:1809.10166 [hep-ph]} \BibitemShut
  {NoStop}%
\bibitem [{\citenamefont {Burrage}\ \emph
  {et~al.}(2019{\natexlab{a}})\citenamefont {Burrage}, \citenamefont
  {K\"ading}, \citenamefont {Millington},\ and\ \citenamefont
  {Min\'a\v{r}}}]{Burrage2018}%
  \BibitemOpen
  \bibfield  {author} {\bibinfo {author} {\bibfnamefont {C.}~\bibnamefont
  {Burrage}}, \bibinfo {author} {\bibfnamefont {C.}~\bibnamefont {K\"ading}},
  \bibinfo {author} {\bibfnamefont {P.}~\bibnamefont {Millington}},\ and\
  \bibinfo {author} {\bibfnamefont {J.}~\bibnamefont {Min\'a\v{r}}},\
  }\bibfield  {title} {\bibinfo {title} {{Open quantum dynamics induced by
  light scalar fields}},\ }\href {https://doi.org/10.1103/PhysRevD.100.076003}
  {\bibfield  {journal} {\bibinfo  {journal} {Phys. Rev. D}\ }\textbf {\bibinfo
  {volume} {100}},\ \bibinfo {pages} {076003} (\bibinfo {year}
  {2019}{\natexlab{a}})},\ \Eprint {https://arxiv.org/abs/1812.08760}
  {arXiv:1812.08760 [hep-th]} \BibitemShut {NoStop}%
\bibitem [{\citenamefont {Burrage}\ \emph
  {et~al.}(2019{\natexlab{b}})\citenamefont {Burrage}, \citenamefont
  {K\"ading}, \citenamefont {Millington},\ and\ \citenamefont
  {Min\'a\v{r}}}]{Burrage2019}%
  \BibitemOpen
  \bibfield  {author} {\bibinfo {author} {\bibfnamefont {C.}~\bibnamefont
  {Burrage}}, \bibinfo {author} {\bibfnamefont {C.}~\bibnamefont {K\"ading}},
  \bibinfo {author} {\bibfnamefont {P.}~\bibnamefont {Millington}},\ and\
  \bibinfo {author} {\bibfnamefont {J.}~\bibnamefont {Min\'a\v{r}}},\
  }\bibfield  {title} {\bibinfo {title} {{Influence functionals, decoherence
  and conformally coupled scalars}},\ }\href
  {https://doi.org/10.1088/1742-6596/1275/1/012041} {\bibfield  {journal}
  {\bibinfo  {journal} {J. Phys. Conf. Ser.}\ }\textbf {\bibinfo {volume}
  {1275}},\ \bibinfo {pages} {012041} (\bibinfo {year} {2019}{\natexlab{b}})},\
  \Eprint {https://arxiv.org/abs/1902.09607} {arXiv:1902.09607 [hep-th]}
  \BibitemShut {NoStop}%
\bibitem [{\citenamefont {K\"ading}(2019)}]{Kading2019}%
  \BibitemOpen
  \bibfield  {author} {\bibinfo {author} {\bibfnamefont {C.}~\bibnamefont
  {K\"ading}},\ }\emph {\bibinfo {title} {{Astro- and Quantum Physical Tests of
  Screened Scalar Fields}}},\ \href@noop {} {Ph.D. thesis},\ \bibinfo  {school}
  {University of Nottingham}, \bibinfo {address} {Nottingham NG7 2RD, UK}
  (\bibinfo {year} {2019}),\ \Eprint {https://arxiv.org/abs/1910.05738}
  {arXiv:1910.05738 [gr-qc]} \BibitemShut {NoStop}%
\bibitem [{\citenamefont {Hartley}\ \emph
  {et~al.}(2019{\natexlab{b}})\citenamefont {Hartley}, \citenamefont
  {K\"ading}, \citenamefont {Howl},\ and\ \citenamefont
  {Fuentes}}]{Hartley2018}%
  \BibitemOpen
  \bibfield  {author} {\bibinfo {author} {\bibfnamefont {D.}~\bibnamefont
  {Hartley}}, \bibinfo {author} {\bibfnamefont {C.}~\bibnamefont {K\"ading}},
  \bibinfo {author} {\bibfnamefont {R.}~\bibnamefont {Howl}},\ and\ \bibinfo
  {author} {\bibfnamefont {I.}~\bibnamefont {Fuentes}},\ }\bibfield  {title}
  {\bibinfo {title} {{Quantum simulation of dark energy candidates}},\ }\href
  {https://doi.org/10.1103/PhysRevD.99.105002} {\bibfield  {journal} {\bibinfo
  {journal} {Phys. Rev. D}\ }\textbf {\bibinfo {volume} {99}},\ \bibinfo
  {pages} {105002} (\bibinfo {year} {2019}{\natexlab{b}})},\ \Eprint
  {https://arxiv.org/abs/1811.06927} {arXiv:1811.06927 [gr-qc]} \BibitemShut
  {NoStop}%
\bibitem [{\citenamefont {Burrage}\ \emph {et~al.}(2017)\citenamefont
  {Burrage}, \citenamefont {Copeland},\ and\ \citenamefont
  {Millington}}]{Burrage2016_2}%
  \BibitemOpen
  \bibfield  {author} {\bibinfo {author} {\bibfnamefont {C.}~\bibnamefont
  {Burrage}}, \bibinfo {author} {\bibfnamefont {E.~J.}\ \bibnamefont
  {Copeland}},\ and\ \bibinfo {author} {\bibfnamefont {P.}~\bibnamefont
  {Millington}},\ }\bibfield  {title} {\bibinfo {title} {{Radial acceleration
  relation from symmetron fifth forces}},\ }\href
  {https://doi.org/10.1103/PhysRevD.95.064050} {\bibfield  {journal} {\bibinfo
  {journal} {Phys. Rev. D}\ }\textbf {\bibinfo {volume} {95}},\ \bibinfo
  {pages} {064050} (\bibinfo {year} {2017})},\ \bibinfo {note} {[Erratum:
  Phys.Rev.D 95, 129902 (2017)]},\ \Eprint {https://arxiv.org/abs/1610.07529}
  {arXiv:1610.07529 [astro-ph.CO]} \BibitemShut {NoStop}%
\bibitem [{\citenamefont {O'Hare}\ and\ \citenamefont
  {Burrage}(2018)}]{OHare:2018ayv}%
  \BibitemOpen
  \bibfield  {author} {\bibinfo {author} {\bibfnamefont {C.~A.~J.}\
  \bibnamefont {O'Hare}}\ and\ \bibinfo {author} {\bibfnamefont
  {C.}~\bibnamefont {Burrage}},\ }\bibfield  {title} {\bibinfo {title}
  {{Stellar kinematics from the symmetron fifth force in the Milky Way disk}},\
  }\href {https://doi.org/10.1103/PhysRevD.98.064019} {\bibfield  {journal}
  {\bibinfo  {journal} {Phys. Rev. D}\ }\textbf {\bibinfo {volume} {98}},\
  \bibinfo {pages} {064019} (\bibinfo {year} {2018})},\ \Eprint
  {https://arxiv.org/abs/1805.05226} {arXiv:1805.05226 [astro-ph.CO]}
  \BibitemShut {NoStop}%
\bibitem [{\citenamefont {Burrage}\ \emph
  {et~al.}(2019{\natexlab{c}})\citenamefont {Burrage}, \citenamefont
  {Copeland}, \citenamefont {K\"ading},\ and\ \citenamefont
  {Millington}}]{Burrage2018Sym}%
  \BibitemOpen
  \bibfield  {author} {\bibinfo {author} {\bibfnamefont {C.}~\bibnamefont
  {Burrage}}, \bibinfo {author} {\bibfnamefont {E.~J.}\ \bibnamefont
  {Copeland}}, \bibinfo {author} {\bibfnamefont {C.}~\bibnamefont {K\"ading}},\
  and\ \bibinfo {author} {\bibfnamefont {P.}~\bibnamefont {Millington}},\
  }\bibfield  {title} {\bibinfo {title} {{Symmetron scalar fields: Modified
  gravity, dark matter, or both?}},\ }\href
  {https://doi.org/10.1103/PhysRevD.99.043539} {\bibfield  {journal} {\bibinfo
  {journal} {Phys. Rev. D}\ }\textbf {\bibinfo {volume} {99}},\ \bibinfo
  {pages} {043539} (\bibinfo {year} {2019}{\natexlab{c}})},\ \Eprint
  {https://arxiv.org/abs/1811.12301} {arXiv:1811.12301 [astro-ph.CO]}
  \BibitemShut {NoStop}%
\bibitem [{\citenamefont {Brax}\ \emph
  {et~al.}(2012{\natexlab{a}})\citenamefont {Brax}, \citenamefont {Davis},\
  and\ \citenamefont {Li}}]{Brax2011}%
  \BibitemOpen
  \bibfield  {author} {\bibinfo {author} {\bibfnamefont {P.}~\bibnamefont
  {Brax}}, \bibinfo {author} {\bibfnamefont {A.-C.}\ \bibnamefont {Davis}},\
  and\ \bibinfo {author} {\bibfnamefont {B.}~\bibnamefont {Li}},\ }\bibfield
  {title} {\bibinfo {title} {{Modified Gravity Tomography}},\ }\href
  {https://doi.org/10.1016/j.physletb.2012.08.002} {\bibfield  {journal}
  {\bibinfo  {journal} {Phys. Lett. B}\ }\textbf {\bibinfo {volume} {715}},\
  \bibinfo {pages} {38} (\bibinfo {year} {2012}{\natexlab{a}})},\ \Eprint
  {https://arxiv.org/abs/1111.6613} {arXiv:1111.6613 [astro-ph.CO]}
  \BibitemShut {NoStop}%
\bibitem [{\citenamefont {Brax}\ \emph
  {et~al.}(2012{\natexlab{b}})\citenamefont {Brax}, \citenamefont {Davis},
  \citenamefont {Li},\ and\ \citenamefont {Winther}}]{Brax2012}%
  \BibitemOpen
  \bibfield  {author} {\bibinfo {author} {\bibfnamefont {P.}~\bibnamefont
  {Brax}}, \bibinfo {author} {\bibfnamefont {A.-C.}\ \bibnamefont {Davis}},
  \bibinfo {author} {\bibfnamefont {B.}~\bibnamefont {Li}},\ and\ \bibinfo
  {author} {\bibfnamefont {H.~A.}\ \bibnamefont {Winther}},\ }\bibfield
  {title} {\bibinfo {title} {{A Unified Description of Screened Modified
  Gravity}},\ }\href {https://doi.org/10.1103/PhysRevD.86.044015} {\bibfield
  {journal} {\bibinfo  {journal} {Phys. Rev. D}\ }\textbf {\bibinfo {volume}
  {86}},\ \bibinfo {pages} {044015} (\bibinfo {year} {2012}{\natexlab{b}})},\
  \Eprint {https://arxiv.org/abs/1203.4812} {arXiv:1203.4812 [astro-ph.CO]}
  \BibitemShut {NoStop}%
\bibitem [{\citenamefont {Brax}\ and\ \citenamefont
  {Burrage}(2014)}]{Brax:2014vva}%
  \BibitemOpen
  \bibfield  {author} {\bibinfo {author} {\bibfnamefont {P.}~\bibnamefont
  {Brax}}\ and\ \bibinfo {author} {\bibfnamefont {C.}~\bibnamefont {Burrage}},\
  }\bibfield  {title} {\bibinfo {title} {{Constraining Disformally Coupled
  Scalar Fields}},\ }\href {https://doi.org/10.1103/PhysRevD.90.104009}
  {\bibfield  {journal} {\bibinfo  {journal} {Phys. Rev. D}\ }\textbf {\bibinfo
  {volume} {90}},\ \bibinfo {pages} {104009} (\bibinfo {year} {2014})},\
  \Eprint {https://arxiv.org/abs/1407.1861} {arXiv:1407.1861 [astro-ph.CO]}
  \BibitemShut {NoStop}%
\bibitem [{\citenamefont {Brax}\ \emph {et~al.}(2015)\citenamefont {Brax},
  \citenamefont {Burrage},\ and\ \citenamefont {Englert}}]{Brax:2015hma}%
  \BibitemOpen
  \bibfield  {author} {\bibinfo {author} {\bibfnamefont {P.}~\bibnamefont
  {Brax}}, \bibinfo {author} {\bibfnamefont {C.}~\bibnamefont {Burrage}},\ and\
  \bibinfo {author} {\bibfnamefont {C.}~\bibnamefont {Englert}},\ }\bibfield
  {title} {\bibinfo {title} {{Disformal dark energy at colliders}},\ }\href
  {https://doi.org/10.1103/PhysRevD.92.044036} {\bibfield  {journal} {\bibinfo
  {journal} {Phys. Rev. D}\ }\textbf {\bibinfo {volume} {92}},\ \bibinfo
  {pages} {044036} (\bibinfo {year} {2015})},\ \Eprint
  {https://arxiv.org/abs/1506.04057} {arXiv:1506.04057 [hep-ph]} \BibitemShut
  {NoStop}%
\bibitem [{\citenamefont {Bekenstein}(1993)}]{Bekenstein:1992pj}%
  \BibitemOpen
  \bibfield  {author} {\bibinfo {author} {\bibfnamefont {J.~D.}\ \bibnamefont
  {Bekenstein}},\ }\bibfield  {title} {\bibinfo {title} {{The Relation between
  physical and gravitational geometry}},\ }\href
  {https://doi.org/10.1103/PhysRevD.48.3641} {\bibfield  {journal} {\bibinfo
  {journal} {Phys. Rev. D}\ }\textbf {\bibinfo {volume} {48}},\ \bibinfo
  {pages} {3641} (\bibinfo {year} {1993})},\ \Eprint
  {https://arxiv.org/abs/gr-qc/9211017} {arXiv:gr-qc/9211017} \BibitemShut
  {NoStop}%
\bibitem [{\citenamefont {Dong}\ \emph {et~al.}(2014)\citenamefont {Dong},
  \citenamefont {Kinney},\ and\ \citenamefont {Stojkovic}}]{Dong:2013swa}%
  \BibitemOpen
  \bibfield  {author} {\bibinfo {author} {\bibfnamefont {R.}~\bibnamefont
  {Dong}}, \bibinfo {author} {\bibfnamefont {W.~H.}\ \bibnamefont {Kinney}},\
  and\ \bibinfo {author} {\bibfnamefont {D.}~\bibnamefont {Stojkovic}},\
  }\bibfield  {title} {\bibinfo {title} {{Symmetron Inflation}},\ }\href
  {https://doi.org/10.1088/1475-7516/2014/01/021} {\bibfield  {journal}
  {\bibinfo  {journal} {JCAP}\ }\textbf {\bibinfo {volume} {01}},\ \bibinfo
  {pages} {021}},\ \Eprint {https://arxiv.org/abs/1307.4451} {arXiv:1307.4451
  [astro-ph.CO]} \BibitemShut {NoStop}%
\bibitem [{\citenamefont {Solomon}\ \emph {et~al.}(2022)\citenamefont
  {Solomon}, \citenamefont {Agarwal},\ and\ \citenamefont
  {Stojkovic}}]{Solomon:2022qqf}%
  \BibitemOpen
  \bibfield  {author} {\bibinfo {author} {\bibfnamefont {R.}~\bibnamefont
  {Solomon}}, \bibinfo {author} {\bibfnamefont {G.}~\bibnamefont {Agarwal}},\
  and\ \bibinfo {author} {\bibfnamefont {D.}~\bibnamefont {Stojkovic}},\
  }\bibfield  {title} {\bibinfo {title} {Environment dependent electron mass
  and the hubble constant tension},\ }\href
  {https://doi.org/10.1103/PhysRevD.105.103536} {\bibfield  {journal} {\bibinfo
   {journal} {Phys. Rev. D}\ }\textbf {\bibinfo {volume} {105}},\ \bibinfo
  {pages} {103536} (\bibinfo {year} {2022})},\ \Eprint
  {https://arxiv.org/abs/2201.03127} {arXiv:2201.03127 [hep-ph]} \BibitemShut
  {NoStop}%
\bibitem [{\citenamefont {Burrage}\ \emph
  {et~al.}(2016{\natexlab{b}})\citenamefont {Burrage}, \citenamefont
  {Copeland},\ and\ \citenamefont {Millington}}]{Burrage2016}%
  \BibitemOpen
  \bibfield  {author} {\bibinfo {author} {\bibfnamefont {C.}~\bibnamefont
  {Burrage}}, \bibinfo {author} {\bibfnamefont {E.~J.}\ \bibnamefont
  {Copeland}},\ and\ \bibinfo {author} {\bibfnamefont {P.}~\bibnamefont
  {Millington}},\ }\bibfield  {title} {\bibinfo {title} {{Radiative Screening
  of Fifth Forces}},\ }\href {https://doi.org/10.1103/PhysRevLett.117.211102}
  {\bibfield  {journal} {\bibinfo  {journal} {Phys. Rev. Lett.}\ }\textbf
  {\bibinfo {volume} {117}},\ \bibinfo {pages} {211102} (\bibinfo {year}
  {2016}{\natexlab{b}})},\ \Eprint {https://arxiv.org/abs/1604.06051}
  {arXiv:1604.06051 [gr-qc]} \BibitemShut {NoStop}%
\bibitem [{\citenamefont {Dyson}\ \emph {et~al.}(1920)\citenamefont {Dyson},
  \citenamefont {Eddington},\ and\ \citenamefont {Davidson}}]{Dyson:1920cwa}%
  \BibitemOpen
  \bibfield  {author} {\bibinfo {author} {\bibfnamefont {F.~W.}\ \bibnamefont
  {Dyson}}, \bibinfo {author} {\bibfnamefont {A.~S.}\ \bibnamefont
  {Eddington}},\ and\ \bibinfo {author} {\bibfnamefont {C.}~\bibnamefont
  {Davidson}},\ }\bibfield  {title} {\bibinfo {title} {{A Determination of the
  Deflection of Light by the Sun's Gravitational Field, from Observations Made
  at the Total Eclipse of May 29, 1919}},\ }\href
  {https://doi.org/10.1098/rsta.1920.0009} {\bibfield  {journal} {\bibinfo
  {journal} {Phil. Trans. Roy. Soc. Lond. A}\ }\textbf {\bibinfo {volume}
  {220}},\ \bibinfo {pages} {291} (\bibinfo {year} {1920})}\BibitemShut
  {NoStop}%
\bibitem [{\citenamefont {Massey}\ \emph {et~al.}(2010)\citenamefont {Massey},
  \citenamefont {Kitching},\ and\ \citenamefont {Richard}}]{Massey:2010hh}%
  \BibitemOpen
  \bibfield  {author} {\bibinfo {author} {\bibfnamefont {R.}~\bibnamefont
  {Massey}}, \bibinfo {author} {\bibfnamefont {T.}~\bibnamefont {Kitching}},\
  and\ \bibinfo {author} {\bibfnamefont {J.}~\bibnamefont {Richard}},\
  }\bibfield  {title} {\bibinfo {title} {{The dark matter of gravitational
  lensing}},\ }\href {https://doi.org/10.1088/0034-4885/73/8/086901} {\bibfield
   {journal} {\bibinfo  {journal} {Rept. Prog. Phys.}\ }\textbf {\bibinfo
  {volume} {73}},\ \bibinfo {pages} {086901} (\bibinfo {year} {2010})},\
  \Eprint {https://arxiv.org/abs/1001.1739} {arXiv:1001.1739 [astro-ph.CO]}
  \BibitemShut {NoStop}%
\bibitem [{\citenamefont {Weinberg}\ \emph {et~al.}(2008)\citenamefont
  {Weinberg}, \citenamefont {Colombi}, \citenamefont {Davé},\ and\
  \citenamefont {Katz}}]{Weinberg_2008}%
  \BibitemOpen
  \bibfield  {author} {\bibinfo {author} {\bibfnamefont {D.~H.}\ \bibnamefont
  {Weinberg}}, \bibinfo {author} {\bibfnamefont {S.}~\bibnamefont {Colombi}},
  \bibinfo {author} {\bibfnamefont {R.}~\bibnamefont {Davé}},\ and\ \bibinfo
  {author} {\bibfnamefont {N.}~\bibnamefont {Katz}},\ }\bibfield  {title}
  {\bibinfo {title} {{Baryon Dynamics, Dark Matter Substructure, and
  Galaxies}},\ }\href {https://doi.org/10.1086/524646} {\bibfield  {journal}
  {\bibinfo  {journal} {The Astrophysical Journal}\ }\textbf {\bibinfo {volume}
  {678}},\ \bibinfo {pages} {6} (\bibinfo {year} {2008})}\BibitemShut {NoStop}%
\bibitem [{\citenamefont {Amendola}\ and\ \citenamefont
  {Tsujikawa}(2015)}]{Amendola:2015ksp}%
  \BibitemOpen
  \bibfield  {author} {\bibinfo {author} {\bibfnamefont {L.}~\bibnamefont
  {Amendola}}\ and\ \bibinfo {author} {\bibfnamefont {S.}~\bibnamefont
  {Tsujikawa}},\ }\href@noop {} {\emph {\bibinfo {title} {{Dark Energy}:
  {Theory and Observations}}}}\ (\bibinfo  {publisher} {Cambridge University
  Press},\ \bibinfo {year} {2015})\BibitemShut {NoStop}%
\bibitem [{\citenamefont {Fathi}\ \emph {et~al.}(2010)\citenamefont {Fathi},
  \citenamefont {Allen}, \citenamefont {Boch}, \citenamefont {Hatziminaoglou},\
  and\ \citenamefont {Peletier}}]{Fathi_2010}%
  \BibitemOpen
  \bibfield  {author} {\bibinfo {author} {\bibfnamefont {K.}~\bibnamefont
  {Fathi}}, \bibinfo {author} {\bibfnamefont {M.}~\bibnamefont {Allen}},
  \bibinfo {author} {\bibfnamefont {T.}~\bibnamefont {Boch}}, \bibinfo {author}
  {\bibfnamefont {E.}~\bibnamefont {Hatziminaoglou}},\ and\ \bibinfo {author}
  {\bibfnamefont {R.~F.}\ \bibnamefont {Peletier}},\ }\bibfield  {title}
  {\bibinfo {title} {Scalelength of disc galaxies},\ }\href
  {https://doi.org/10.1111/j.1365-2966.2010.16812.x} {\bibfield  {journal}
  {\bibinfo  {journal} {Monthly Notices of the Royal Astronomical Society}\ ,\
  \bibinfo {pages} {no}} (\bibinfo {year} {2010})}\BibitemShut {NoStop}%
\bibitem [{\citenamefont {Prat}\ \emph {et~al.}(2018)\citenamefont {Prat} \emph
  {et~al.}}]{DES:2017gwu}%
  \BibitemOpen
  \bibfield  {author} {\bibinfo {author} {\bibfnamefont {J.}~\bibnamefont
  {Prat}} \emph {et~al.} (\bibinfo {collaboration} {DES}),\ }\bibfield  {title}
  {\bibinfo {title} {{Dark Energy Survey year 1 results: Galaxy-galaxy
  lensing}},\ }\href {https://doi.org/10.1103/PhysRevD.98.042005} {\bibfield
  {journal} {\bibinfo  {journal} {Phys. Rev. D}\ }\textbf {\bibinfo {volume}
  {98}},\ \bibinfo {pages} {042005} (\bibinfo {year} {2018})},\ \Eprint
  {https://arxiv.org/abs/1708.01537} {arXiv:1708.01537 [astro-ph.CO]}
  \BibitemShut {NoStop}%
\bibitem [{\citenamefont {Hogg}(1999)}]{Hogg:1999ad}%
  \BibitemOpen
  \bibfield  {author} {\bibinfo {author} {\bibfnamefont {D.~W.}\ \bibnamefont
  {Hogg}},\ }\bibfield  {title} {\bibinfo {title} {{Distance measures in
  cosmology}},\ }\href@noop {} {\  (\bibinfo {year} {1999})},\ \Eprint
  {https://arxiv.org/abs/astro-ph/9905116} {arXiv:astro-ph/9905116}
  \BibitemShut {NoStop}%
\bibitem [{\citenamefont {Prat}\ \emph {et~al.}(2022)\citenamefont {Prat},
  \citenamefont {Hogan}, \citenamefont {Chang},\ and\ \citenamefont
  {Frieman}}]{Prat:2021xlz}%
  \BibitemOpen
  \bibfield  {author} {\bibinfo {author} {\bibfnamefont {J.}~\bibnamefont
  {Prat}}, \bibinfo {author} {\bibfnamefont {C.}~\bibnamefont {Hogan}},
  \bibinfo {author} {\bibfnamefont {C.}~\bibnamefont {Chang}},\ and\ \bibinfo
  {author} {\bibfnamefont {J.}~\bibnamefont {Frieman}},\ }\bibfield  {title}
  {\bibinfo {title} {{Vacuum energy density measured from cosmological data}},\
  }\href {https://doi.org/10.1088/1475-7516/2022/06/015} {\bibfield  {journal}
  {\bibinfo  {journal} {JCAP}\ }\textbf {\bibinfo {volume} {06}}\bibfield
  {number} {\bibinfo  {number} { (06)},\ \bibinfo {pages} {015}},\ }\Eprint
  {https://arxiv.org/abs/2111.08151} {arXiv:2111.08151 [astro-ph.CO]}
  \BibitemShut {NoStop}%
\bibitem [{\citenamefont {Brax}\ and\ \citenamefont
  {Davis}(2016)}]{Brax:2016wjk}%
  \BibitemOpen
  \bibfield  {author} {\bibinfo {author} {\bibfnamefont {P.}~\bibnamefont
  {Brax}}\ and\ \bibinfo {author} {\bibfnamefont {A.-C.}\ \bibnamefont
  {Davis}},\ }\bibfield  {title} {\bibinfo {title} {{Atomic Interferometry Test
  of Dark Energy}},\ }\href {https://doi.org/10.1103/PhysRevD.94.104069}
  {\bibfield  {journal} {\bibinfo  {journal} {Phys. Rev. D}\ }\textbf {\bibinfo
  {volume} {94}},\ \bibinfo {pages} {104069} (\bibinfo {year} {2016})},\
  \Eprint {https://arxiv.org/abs/1609.09242} {arXiv:1609.09242 [astro-ph.CO]}
  \BibitemShut {NoStop}%
\end{thebibliography}%

\end{document}